\documentclass[aps,prb,reprint,groupedaddress]{revtex4-2}
\usepackage{physics}
\usepackage{siunitx}
\usepackage{mathtools}
\usepackage{amssymb}
\usepackage{bm}      
\usepackage{graphicx}
\graphicspath{{./Figures/}}
\usepackage{hyperref}
\hypersetup{unicode=true, colorlinks=true, linkcolor=blue, citecolor=blue}
\usepackage{titlesec}
\titlespacing*{\subsection}{0pt}{*3}{0pt} 

\usepackage{enumitem}
\usepackage{textcase}
\makeatletter

\def\rtx@sectionbase{%
  \@startsection{section}{1}{\z@}%
    {1\baselineskip}
    {1\baselineskip}
    {\normalfont\bfseries\centering}
}

\renewcommand{\section}{\@ifstar{\rtx@sectstar}{\rtx@sect}}
\newcommand{\rtx@sect}[2][]{%
  \rtx@sectionbase[\MakeTextUppercase{#1}]{\MakeTextUppercase{#2}}%
}
\newcommand{\rtx@sectstar}[1]{%
  \rtx@sectionbase*{\MakeTextUppercase{#1}}%
}

\makeatother

\begin{document}

\title{Quantum Tomography of Suspended Carbon Nanotubes}

\author{Jialiang Chang}%
\affiliation{Department of Physics and Astronomy, Tufts University, Medford, MA, 02155, USA}

\author{Nicholas Pietrzak}%
\affiliation{Department of Physics and Astronomy, Tufts University, Medford, MA, 02155, USA}

\author{Cristian Staii}%
\email{cstaii01@tufts.edu}
\affiliation{Department of Physics and Astronomy, Tufts University, Medford, MA, 02155, USA}

\date{\today}

\begin{abstract}

We propose and analyze an all-mechanical route to coherent control and quantum-state reconstruction of the fundamental flexural mode of a suspended carbon nanotube (CNT) operated in the anharmonic (Duffing/Kerr) regime. A nearby atomic force microscope (AFM) tip provides a single, localized actuator that applies calibrated, time-dependent forces to the CNT. In the presence of mechanical anharmonicity this enables spectrally selective control of the lowest vibrational transition and thus supports effective two-levels protocols such as Rabi oscillations and Ramsey interferometry. The same actuator also implements the controlled phase-space displacements required for Wigner function tomography via displaced-parity sampling, thereby unifying control and tomography without optical heating and without dedicated on-chip microwave drive lines at the CNT resonator. We develop explicit pulse sequences and a master equation framework that connect experimentally accessible signals to energy relaxation and phase coherence times and to parity-based quantum signatures, including negative regions of the Wigner function. The approach is compatible with multiple readout modalities, including direct AFM-based detection and dispersive coupling to superconducting circuitry such as a Cooper-pair box and/or a microwave cavity. Together, these techniques provide complete access to populations, coherence, and parity within a single device architecture. This minimal scheme provides a practical route to all-mechanical quantum control and state-resolved characterization of decoherence in mesoscopic mechanical systems.

\end{abstract}

\maketitle

\section{Introduction}

Probing quantum decoherence in mesoscopic mechanical systems remains a central challenge for both emerging quantum technologies and experimental tests of macroscopic quantum phenomena \cite{armour_entanglement_2002,lee_strong_2023,schwab_putting_2005,blencowe_quantum_2004}. Nanomechanical resonators are particularly attractive in this context because they can be engineered with long lifetimes, couple sensitively to their environment, and admit dispersive measurement strategies that access motional observables at the quantum level \cite{samanta_nonlinear_2023,lahaye_approaching_2004,remus_damping_2009,wollack_quantum_2022,schlosshauer_decoherence_2007}. Over the past two decades, the key experimental ingredients required to reach and operate in the quantum regime have been established across multiple platforms, including cryogenic operation, feedback or sideband cooling, high-$Q$ suspension, and quantum-limited readout. Representative milestones include ground-state cooling \cite{teufel_sideband_2011}, preparation and control at the single-quantum level \cite{oconnell_quantum_2010}, ultralong phonon lifetimes in engineered acoustic structures \cite{maccabe_nano-acoustic_2020}, and measurement-based quantum control of mechanical motion \cite{rossi_measurement-based_2018}.

Within these platforms, suspended carbon nanotubes (CNTs) stand out as particularly compelling mechanical resonators: they combine extremely small motional masses with high frequencies and large zero-point motion, and their mechanical properties can be tuned \emph{in situ} via electrostatic gating and tension control \cite{sazonova_tunable_2004,garcia-sanchez_mechanical_2007,huttel_nanoelectromechanics_2008,tavernarakis_optomechanics_2018}. These attributes enable strong coupling to localized forces and make CNTs natural candidates for studies of decoherence and nonequilibrium quantum dynamics in nanomechanics \cite{schneider_observation_2014,palyi_spinorbitinduced_2012,wang_hybrid_2017,wang_creating_2016}. At the same time, CNT resonators are intrinsically susceptible to nonlinear effects. Geometric nonlinearity of a doubly clamped beam and electrostatic nonlinearities from nearby electrodes generically produce Duffing-type dynamics in the classical regime and the corresponding Kerr-type anharmonicity in the quantized description \cite{semiao_kerr_2009, lifshitz_cross_nonlinear_2009,rhoads_nonlinear_2010,eichler_nonlinear_2011,rips_nonlinear_2014,samanta_nonlinear_2023}. This nonlinearity is not merely a complication: when the anharmonicity is sufficiently large compared to the mechanical linewidth and the relevant control bandwidth, it enables spectrally selective addressing of the lowest transition and thereby supports effective two-level control protocols in a mechanical degree of freedom \cite{rips_nonlinear_2014,wang_method_2016,sarma_tunable_2018}.

A recurring experimental challenge in mesoscopic nanomechanics is that actuation, coherent control, and quantum-state reconstruction are often implemented through distinct physical subsystems, for example optical actuation combined with microwave dispersive readout, or on-chip microwave control paired with a separate tomographic procedure \cite{oconnell_quantum_2010,rossi_measurement-based_2018,teufel_sideband_2011,lahaye_approaching_2004}. This partitioning can increase device complexity, introduce additional dissipation channels, and complicate a clean interpretation of decoherence due to the environment of the mechanical mode itself \cite{lee_strong_2023,remus_damping_2009,schlosshauer_decoherence_2007}. In CNT devices, these issues are amplified by the need to maintain ultralow heating and to preserve high-$Q$ operation while retaining local, well-calibrated control over the mechanical degree of freedom.

In this paper we propose and analyze a compact protocol for coherent control and quantum-state reconstruction of a single suspended CNT operating in the anharmonic (Duffing/Kerr) regime. The central idea is to use a nearby atomic force microscope (AFM) tip as a single, precisely positioned actuator that applies calibrated forces to the nanotube through modulation of the tip--CNT interaction. Importantly, the actuation bandwidth is set by electrical modulation of the tip--CNT interaction, while the AFM cantilever and positioning stack provide quasi-static alignment and do not need to resonate at that frequency \cite{garcia-sanchez_mechanical_2007,garcia-sanchez_imaging_2008}. This separation allows one to combine nanoscale spatial selectivity with resonant driving and readout of the CNT mode under dilution-refrigerator conditions. The protocol naturally separates two operating modes that are standard in driven resonator experiments: continuous-wave (CW) excitation for spectroscopy and calibration of the driven response, and time-domain (envelope-shaped) control segments used for Ramsey interferometry and phase-space displacements. This connection to classical forced-vibration response provides a practical route for relating applied drive waveforms to experimentally measured amplitudes, linewidths, and the control parameters entering the quantum model \cite{lifshitz_cross_nonlinear_2009,rhoads_nonlinear_2010}.

On the quantum side, the CNT anharmonicity enables selective addressing of the lowest $|0\rangle\leftrightarrow|1\rangle$ transition over the relevant spectral widths, thereby supporting time-domain control protocols analogous to Rabi and Ramsey measurements within an effective two-level manifold. For state reconstruction, the same actuator provides controlled displacements in phase space that can be combined with a parity-sensitive readout to access the Wigner function. The proposed architecture is compatible with several complementary transduction pathways, including direct AFM-based detection and dispersive coupling to superconducting circuitry such as a Cooper-pair box (CPB) and/or a microwave cavity \cite{armour_entanglement_2002, blencowe_quantum_2004, pirkkalainen_hybrid_2013, blais_circuit_2021}. Throughout, we emphasize experimentally relevant constraints that determine feasibility in realistic devices, including thermal occupancy and initialization, mechanical linewidth and noise broadening, and force-noise contributions associated with the tip--sample environment.

The novelty of the present proposal is twofold. First, it leverages intrinsic and tunable nonlinearity of a suspended CNT to identify and quantify an operating regime in which the lowest transition can be isolated and treated as an effective two-level system (TLS) \cite{eichler_nonlinear_2011,rips_nonlinear_2014,samanta_nonlinear_2023, semiao_kerr_2009}. Second, it integrates this TLS (qubit-like) control with phase-space displacements produced by the \emph{same} localized actuator, providing a unified route to both decoherence characterization and quantum-state tomography within a single, mechanically addressable device. Establishing such a unified toolbox in CNT nanomechanics would open a practical pathway toward quantitative tests of decoherence mechanisms in a mesoscopic mechanical system and toward ultrasensitive force and field sensing protocols that exploit nonclassical motional states.

This paper is organized as follows: \hyperref[sec:hamiltonian]{Section~II} introduces the device concept and effective Hamiltonian in the anharmonic two-level operating regime. \hyperref[sec:lindblad]{Section~III} presents the Gorini--Kossakowski--Sudarshan--Lindblad master equation and the associated Bloch dynamics. \hyperref[sec:ramsey]{Section~IV} presents the Rabi and Ramsey-interferometry measurements. \hyperref[sec:wigner]{Section~V} develops the Wigner-tomography protocols, including displaced-parity reconstruction, and discusses readout modalities. \hyperref[sec:estimates]{Section~VI} provides numerical estimates and parameter ranges relevant to feasibility and to the validity of the underlying approximations, and discusses the AFM--CNT interaction. \hyperref[sec:conclusions]{Section~VII} summarizes the main results.

\section{Effective Two-Level Model and Hamiltonian}\label{sec:hamiltonian}
\subsection{Anharmonic CNT mode and controlled reduction to an effective two-level subspace}

\begin{figure}[ht] \includegraphics[scale=0.21]{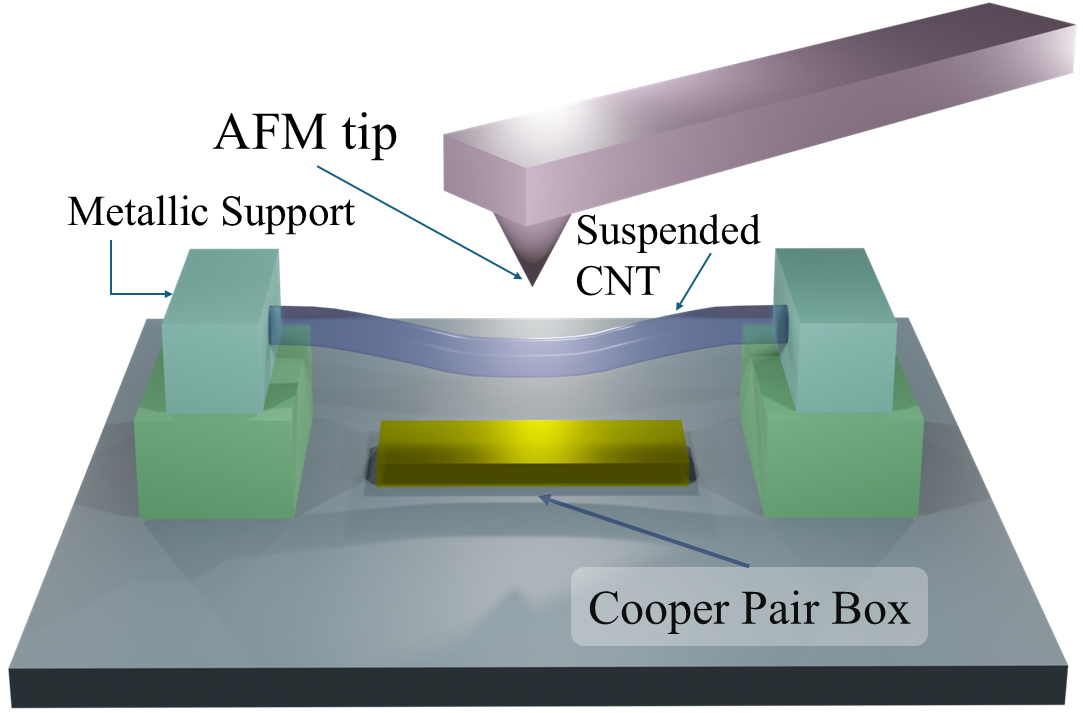} \caption{\label{fig:schematics}Schematic of the proposed experimental setup. A doubly clamped single-walled carbon nanotube (CNT) of length \(L=100\text{--}1000\,\mathrm{nm}\) is suspended over a \(200\,\mathrm{nm}\) trench and driven by localized forces from a nearby atomic force microscope (AFM) tip. This experimental configuration provides a fully mechanical platform for implementing Ramsey interferometry and Wigner-function tomography of a nanomechanical mode under dilution refrigerator conditions (\(T\sim 10\,\mathrm{mK}\), \(P<10^{-7}\,\mathrm{Torr}\)). Also indicated schematically is a Cooper-pair box patterned on the \(\mathrm{SiO_2}/\mathrm{Si}\) substrate, illustrating one possible dispersive-coupling readout scheme for the CNT.} \end{figure}

\textit{Physical system ---}We consider a single-walled CNT of length $L$ doubly clamped across a nanofabricated trench and positioned near an AFM tip (Fig.~\ref{fig:schematics}). The CNT fundamental flexural mode is a high-$Q$ nanomechanical degree of freedom that can be precooled and further cooled by sideband or feedback techniques to a low mean occupancy, so that its dynamics can be treated quantum mechanically \cite{schwab_putting_2005, lahaye_approaching_2004, WilsonRae2007PRL, Urgell2020NatPhys, oconnell_quantum_2010, maccabe_nano-acoustic_2020, rossi_measurement-based_2018, pirkkalainen_hybrid_2013, blais_circuit_2021}.

A linearly driven harmonic mechanical mode is naturally described in terms of phase-space displacements, with dynamics that populate a ladder of number states. To access two-level control protocols in a mechanical degree of freedom, it is therefore essential to operate in a regime where the vibrational spectrum is appreciably \emph{anharmonic}, so that the lowest transition can be addressed selectively while higher transitions are spectrally detuned. Such anharmonicity is well motivated for doubly clamped nanomechanical resonators: flexural motion generically induces geometric nonlinearity through stretching under transverse deflection, which appears as a Duffing-type nonlinearity in the classical description and as an effective quartic correction in the quantized Hamiltonian. This intrinsic mechanism and its quantized counterpart are standard in the nanomechanics literature and form the basis for ``nanomechanical qubit'' proposals that exploit the two lowest vibrational levels of a sufficiently anharmonic resonator \cite{schwab_putting_2005,lahaye_approaching_2004,WilsonRae2007PRL,Urgell2020NatPhys,rips_hartmann_prl_2013, sarma_tunable_2018}. For example, Rips and Hartmann model a doubly clamped nanobeam as a harmonic mode supplemented by a stretching-induced quartic nonlinearity and show how this yields an effective two-level manifold under appropriate operating conditions \cite{rips_hartmann_prl_2013}.

\textit{Enhancing the anharmonicity ---}While intrinsic anharmonicity exists even in the absence of external fields, it can be enhanced and tuned by electrostatic gradient forces from nearby electrodes (or a voltage-biased AFM tip). The key physical point is that suitable electrostatic potentials can soften the mechanical mode and increase the zero-point motion amplitude, thereby amplifying the effective nonlinearity per phonon \cite{rips_nonlinear_2014, lifshitz_cross_nonlinear_2009, huttel_nanoelectromechanics_2008, sazonova_tunable_2004, garcia-sanchez_mechanical_2007, eichler_nonlinear_2011, rhoads_nonlinear_2010}. This, in turn, enhances the anharmonicity and enables selective addressing of an individual transition \cite{rips_hartmann_prl_2013}. In addition, Kerr-type nonlinearities can be engineered in hybrid electromechanical architectures through dispersive coupling between a nanomechanical mode and a superconducting qubit, providing a mechanism for phonon number-dependent phase evolution of the mechanical mode \cite{zueco_qubit_2009, semiao_kerr_2009, wang_method_2016, sarma_tunable_2018}. In our device concept, such engineered nonlinearity can complement the CNT's intrinsic geometric nonlinearity when a superconducting circuit readout is employed, as discussed in \hyperref[sec:wigner]{Section~V}.

Motivated by these established mechanisms, we model the CNT fundamental mode as an anharmonic oscillator described by \cite{SM}
\begin{equation}
H_{\mathrm{m}}=\hbar\omega_{0}\,a^{\dagger}a+\frac{\hbar K}{2}\,a^{\dagger}a^{\dagger}aa,
\label{eq:Hm_anharmonic}
\end{equation}
where $\omega_o$ is the frequency of the fundamental mode and $K$ is the (self-)Kerr anharmonicity (the quantum counterpart of Duffing nonlinearity) \cite{rips_hartmann_prl_2013,lifshitz_cross_nonlinear_2009,rhoads_nonlinear_2010, semiao_kerr_2009}. In this effective description, a quartic correction to the restoring potential of a doubly clamped resonator is reduced (after transformation to the mode basis and within a rotating-wave treatment of the weakly nonlinear regime) to the phonon-number nonlinearity in Eq.~\eqref{eq:Hm_anharmonic} (details in Sec.~S1 of the Supplemental Material \cite{SM} (see also references \cite{blais_cavity_2004, boissonneault_dispersive_2009, cui_feedback_2013, riste_initialization_2012, motzoi_simple_2009}).

\textit{Controlled reduction to an effective two-level subspace--} Let $|n\rangle$ denote the eigenstates of $H_{\mathrm{m}}$. The anharmonicity makes the transition frequencies nonuniform: the $|0\rangle \!\rightarrow\!|1\rangle$ transition differs from $|1\rangle \!\rightarrow\!|2\rangle$ by an amount set by $K$. When this separation is large compared to the relevant spectral widths (mechanical linewidth and drive-induced broadening), the system can be operated as an effective two-level system (TLS) formed by $\{|0\rangle,|1\rangle\}$, with $\omega_{01}$ denoting the transition frequency between the two levels (for the Kerr model of Eq.~\eqref{eq:Hm_anharmonic}, $\omega_{01}=\omega_0$). This is precisely the criterion emphasized in the nanomechanical-qubit literature: provided (i) the motion is initialized near the ground state and (ii) the nonlinearity per phonon is sufficiently large so that higher transitions are off-resonant, dynamics during gate sequences remains confined to $\{|0\rangle,|1\rangle\}$ with high accuracy \cite{rips_hartmann_prl_2013, rips_nonlinear_2014, wang_method_2016, sarma_tunable_2018}. 

A convenient set of sufficient conditions for confinement to the lowest two levels is that the anharmonic separation $|K|$ exceeds both: (1) the homogeneous transition linewidth set by transverse decoherence and (2) the characteristic control bandwidth set by the applied drive \cite{rips_hartmann_prl_2013,wang_method_2016,sarma_tunable_2018}, namely
\begin{equation}
|K| \gg \Gamma_2,\qquad |K| \gg \Omega_R,
\label{eq:tls_conditions}
\end{equation}
where $\Gamma_2$ is the transverse decoherence rate defined in \hyperref[sec:lindblad]{Section~III} and $\Omega_R$ is the on-resonance Rabi frequency generated by the AFM actuation (see below). 
Condition~\eqref{eq:tls_conditions} ensures that a drive tuned to the $|0\rangle\!\rightarrow\!|1\rangle$ transition does not efficiently excite the off-resonant $|1\rangle\!\rightarrow\!|2\rangle$ transition, and that dephasing or spectral diffusion does not wash out the anharmonic selectivity required for two-level operation. In \hyperref[sec:estimates]{Section~VI} we provide numerical estimates of $K$, $\Gamma_2$, and the attainable $\Omega_R$ for the device parameters considered here, thereby delineating the regime in which Eq.~\eqref{eq:tls_conditions} is satisfied.
 Under these conditions we define Pauli operators on the $\{|0\rangle,|1\rangle\}$ manifold,
\begin{equation}
\sigma_z=|1\rangle\!\langle1|-|0\rangle\!\langle0|,\qquad
\sigma_x=|0\rangle\!\langle1|+|1\rangle\!\langle0|,
\end{equation}
and treat the CNT mode as an effective mechanical qubit.

\subsection{AFM actuation and effective two-level system Hamiltonian}

\textit{AFM actuation as a force drive ---} The AFM tip serves both as a nanometer-precision positioning element and as a localized actuator that applies a controlled, time-dependent force to the CNT through modulation of the tip--CNT interaction (for example via an electrostatic force gradient, with possible additional contributions from van der Waals forces). In practice, the high-frequency component of the actuation could be provided by electrically modulating the tip bias (or a local gate potential) at the CNT resonant frequency, while the cantilever itself provides quasi-static positioning and does not need to oscillate at that frequency. This separation of functions, namely low-frequency mechanical positioning combined with high-frequency electrical force modulation, has already been implemented experimentally to drive and characterize suspended CNT and graphene resonators \cite{garcia-sanchez_mechanical_2007, garcia-sanchez_imaging_2008}. An alternative method employs multifrequency AFM to simultaneously record the first eigenmode of the microscope cantilever for topography imaging and the second eigenmode for the detection of the resonator vibration which can be extended in the GHz regime \cite{san-paulo_detection_2007, garcia_emergence_2012}. 

\textit{Tip-position noise and its impact on coherence---} Although the cantilever is not required to oscillate at the CNT resonance, the AFM head, cantilever, and positioning stack constitute a mechanical system coupled to a thermal bath and therefore exhibit low-frequency fluctuations in the tip--CNT separation. In \hyperref[sec:estimates]{Section~VI} we estimate the corresponding noise scales using representative tip--CNT force gradients and experimentally realistic displacement noise levels, and we identify the parameter regime in which these setup-induced contributions remain small compared to the intrinsic linewidth and the anharmonic selectivity required for two-level operation.

\textit{Effective two-level Hamiltonian ---} For continuous-wave (CW) control we model the applied force as $F(t)=F_0\cos(\omega_d t)$ coupling linearly to the CNT displacement operator $\hat{x}$. In a harmonic-mode normalization one may write $\hat{x}=x_{\mathrm{zpf}}(a+a^{\dagger})$, where $x_{\mathrm{zpf}}$ is the zero-point amplitude of the corresponding linearized mode. In the effective two-level manifold, the relevant coupling is set by the transition matrix element $x_{01}\equiv\langle 0|\hat{x}|1\rangle$, so that $\hat{x}\rightarrow x_{01}\sigma_x$. After projecting to $\{|0\rangle,|1\rangle\}$, the driven CNT dynamics is described by
\begin{equation}
H_{\mathrm{TLS}}(t)=\frac{\hbar\omega_{01}}{2}\,\sigma_z
+\hbar g\,\cos(\omega_d t)\,\sigma_x,
\label{eq:H_TLS_lab}
\end{equation}
where $\omega_{01}$ is the anharmonic $|0\rangle \!\rightarrow\!|1\rangle$ transition frequency and $g\equiv F_0 x_{01}/\hbar$ is the drive strength set by the AFM force amplitude $F_0$ and the matrix element $x_{01}$. In the Kerr model of Eq.~\eqref{eq:Hm_anharmonic} the fundamental transition satisfies $\omega_{01}=\omega_0$. For notational simplicity, we therefore write $\omega_0$ for the $\ket{0}\!\leftrightarrow\!\ket{1}$ transition frequency throughout the rest of the paper.  For pulsed control (Ramsey and tomography sequences) $F(t)$ is replaced by an envelope-modulated drive discussed in \hyperref[sec:ramsey]{Sections~IV}. Moving to a frame rotating at $\omega_d$ and invoking the rotating-wave approximation (RWA) reduces Eq.~\eqref{eq:H_TLS_lab} to the standard form

\begin{equation}
H_{\mathrm{TLS}}^{(\mathrm{rot})}=\frac{\hbar\Delta}{2}\,\sigma_z
+\frac{\hbar\Omega_R}{2}\,\sigma_x,
\label{eq:H_TLS_rot}
\end{equation}
where $\Delta\equiv\omega_{0}-\omega_d$ is the detuning and $\Omega_R$ is the on-resonance Rabi frequency set by the drive amplitude (explicit expressions and the derivation are given in Sec.~S1 in the Supplemental Material \cite{SM}). 

We emphasize that the validity of the TLS model is not assumed \emph{a priori}: it is an experimentally testable operating regime. The same spectroscopy used to identify $\omega_{01}$ can also resolve the anharmonic separation between adjacent transitions, and the absence of observable leakage (e.g., no measurable population of higher levels during driven sequences) provides a direct diagnostic that the conditions \eqref{eq:tls_conditions} are satisfied, as in established nanomechanical-qubit protocols \cite{rips_hartmann_prl_2013}. The crucial point is that the anharmonicity renders the two-level truncation quantitatively controlled and allows $\Omega_R$ to be interpreted as the rotation rate of the $|0\rangle \!\leftrightarrow\!|1\rangle$ transition, rather than as the rate of phase-space displacement of a purely harmonic mode.

\subsection{Environmental bath and rotating-frame Hamiltonian}

The CNT mode is coupled to environmental degrees of freedom (clamping loss, charge noise, and other fluctuations) that produce the dominant dephasing channel. In the effective TLS description, these effects can be captured by a bosonic bath and a dominant longitudinal coupling channel (frequency noise) of the form \cite{breuer_theory_2009}
\begin{equation}
H_{B}=\sum_k \hbar\omega_k\,b_k^\dagger b_k,\qquad
H_{I}=\hbar\sigma_z\sum_k g_k\,(b_k^\dagger+b_k),
\label{eq:H_bath_TLS}
\end{equation}
together with the corresponding dissipative dynamics treated in the Born--Markov/Lindblad framework in \hyperref[sec:lindblad]{Section~III}. The coefficients $g_{k}$ in Eq. \eqref{eq:H_bath_TLS} represent coupling to bath mode $k$ with frequency $\omega_{k}$. This effective description is standard for qubit-level decoherence modeling and is appropriate here because the experimentally extracted observables (Ramsey fringe decay, population ringdown) directly determine energy relaxation and phase coherence times within the same reduced two-level manifold (see \hyperref[sec:lindblad]{Section~III}).
Collecting the system, drive, and bath terms we arrive at the rotating‑frame Hamiltonian \cite{SM}:
\begin{equation}
\tilde H = \frac{\hbar\Delta}{2}\,\sigma_{z}
 + \frac{\hbar\Omega_{R}}{2}\,\sigma_{x}
 + \sum_{k}\hbar\omega_{k} b_{k}^{\dagger} b_{k}
 + \hbar \sigma_{z} \sum_{k} g_{k}\bigl(b_{k}^{\dagger}+b_{k}\bigr).
\label{eq:Hrot}
\end{equation}

\textit{Physical picture --} Because the AFM force couples linearly to the displacement operator,
and only the co‑rotating component survives in the RWA, the AFM drive sets a coupling scale of order \(F_{0}x_{01}\), and hence a natural frequency scale $\Omega_{R} \sim ({F_{0}\,x_{01}})/{\hbar}$ for coherent rotations in the effective two-level manifold. When the drive is \emph{on resonance} (\(\Delta=0\)), the nanotube oscillates between
      \(\ket{0}\) and \(\ket{1}\) at the natural frequency scale given by the Rabi frequency
      \(\Omega_{R}\).
Off resonance (\(\Delta\neq0\)) the oscillation frequency generalizes to
      \(\Omega=\sqrt{\Delta^{2}+\Omega_R^{2}}\). 
 
 The bath coupling (last term in Eq.~\eqref{eq:Hrot}) primarily induces dephasing through fluctuations of the transition frequency and thus contributes to the phase-coherence rate $\Gamma_2$. Energy relaxation at rate $\Gamma_1$ is set by mechanical damping and is incorporated in the Lindblad description in \hyperref[sec:lindblad]{Section~III}. Eq. \eqref{eq:Hrot} is the starting point for
analyzing Ramsey and Wigner‑tomography protocols for the
mechanically driven CNT resonator.

\section{Lindblad Master Equation and Bloch Dynamics}\label{sec:lindblad}
 \textit{GKSL equation --} We treat the AFM-driven CNT in the effective two-level manifold as an open quantum system and adopt a Markovian master-equation approach in the rotating frame (see Sec.~S2 in the Supplemental Material for the derivation ~\cite{SM}). Under weak system-bath coupling to a stationary environment with short correlation time (Born--Markov approximation) and within the rotating-wave approximation, the reduced density matrix $\rho_S$ obeys the Gorini--Kossakowski--Sudarshan--Lindblad (GKSL) equation \cite{breuer_theory_2009, schlosshauer_decoherence_2007}: 
\begin{equation}
    \dot{\rho_S} = -\frac{i}{\hbar}[H_S,\rho_S]
    +\gamma_{\downarrow}\,\mathcal{D}[\sigma_-]\rho_S
    +\gamma_{\uparrow}\,\mathcal{D}[\sigma_+]\rho_S
    +\gamma_{\varphi}\,\mathcal{D}[\sigma_z]\rho_S,
    \label{eq:Lindblad}
\end{equation}
where $H_S\equiv H_{\mathrm{TLS}}^{(\mathrm{rot})}$ is given in Eq.~\eqref{eq:H_TLS_rot}, $\sigma_+=\ket{1}\!\bra{0}$, $\sigma_-=\ket{0}\!\bra{1}$, and the \textit{dissipator} is defined as:
\begin{equation}
\mathcal{D}[\sigma]\rho_S\equiv \sigma\rho_S \sigma^{\dagger}
                      -\tfrac12\{\sigma^{\dagger}\sigma,\rho_S\}.
\label{eq:Dissipator}
\end{equation}
Here $\gamma_{\downarrow}$ and $\gamma_{\uparrow}$ describe relaxation ($\ket{1}\to\ket{0}$) and thermal excitation ($\ket{0}\to\ket{1}$), respectively, while $\gamma_{\varphi}$ accounts for pure dephasing from slow longitudinal frequency noise.

\textit{Physical interpretation of dissipators ---} The jump operators $\{\sigma_-,\sigma_+,\sigma_z\}$ encode three distinct channels: (i) energy relaxation at rate $\gamma_{\downarrow}$, (ii) incoherent excitation at rate $\gamma_{\uparrow}$, and (iii) pure dephasing at rate $\gamma_{\varphi}$ that leaves populations unchanged. These processes reduce the visibility of Rabi oscillations and Ramsey fringes and set the time scales relevant for state reconstruction.

\textit{Microscopic rates (model example) ---} For concreteness, and following standard weak-coupling treatments, the excitation and relaxation rates for a transition at angular frequency $\omega_{0}$ can be written in terms of a bath spectral density $J(\omega)$ and the Bose factor $n(\omega)$ as \cite{breuer_theory_2009, schlosshauer_decoherence_2007}:
\begin{equation}
\gamma_{\uparrow}=2\pi J(\omega_{0})\,n(\omega_{0}),\qquad
\gamma_{\downarrow}=2\pi J(\omega_{0})\,[n(\omega_{0})+1],
\label{eq:gamma}
\end{equation}
with $n(\omega)=\bigl[e^{\hbar\omega/k_BT}-1\bigr]^{-1}$, and $J(\omega)=2\alpha\,\omega\,e^{-\omega/\omega_c}$ (Ohmic form with dimensionless coupling $\alpha$ and cutoff $\omega_c$).

\textit{Bloch dynamics for the AFM-driven CNT under dissipation --}  We start with the GKSL equation Eq.~\eqref{eq:Lindblad} and use it to obtain the Bloch-equation dynamics and the relaxation and dephasing rates that enter the control and tomography sequences. The Bloch decay parameters and the corresponding time constants are defined as:
\begin{equation}
T_1\equiv\frac{1}{\Gamma_1},\qquad T_2\equiv\frac{1}{\Gamma_2},\qquad \Gamma_1\equiv\gamma_{\downarrow}+\gamma_{\uparrow},\quad \Gamma_2\equiv\frac{\Gamma_1}{2}+2\gamma_{\varphi},
\label{eq:T1T2}
\end{equation}
The corresponding time constants quantify longitudinal energy relaxation ($T_1$) and transverse phase coherence ($T_2$) \cite{schlosshauer_decoherence_2007, breuer_theory_2009}. A pure dephasing rate arises from slow longitudinal frequency noise and, for an Ohmic bath, scales as $\gamma_{\varphi}\;\simeq\; 2\pi\alpha\,k_BT/\hbar$ \cite{breuer_theory_2009, schlosshauer_decoherence_2007}. Therefore, in the relaxation-limited regime (i.e., when slow longitudinal fluctuations are negligible on the experimental timescale), we have $\gamma_{\varphi}\approx 0$ and thus $\Gamma_{2}=\Gamma_{1}/2$. 

By introducing the Bloch vector components:  
\begin{equation}
 \,X \equiv \Tr(\sigma_{x}\rho_S), \qquad
        Y \equiv \Tr(\sigma_{y}\rho_S), \qquad
        Z \equiv \Tr(\sigma_{z}\rho_S)\,   
\end{equation}
the expression for the reduced density matrix becomes \cite{breuer_theory_2009}:
\begin{equation}
\rho_S=\tfrac12\bigl(\openone+X\sigma_{x}+Y\sigma_{y}+Z\sigma_{z}\bigr).
\label{eq:Matrix}
\end{equation}
Substituting Eq.~\eqref{eq:Matrix} into Eq.~\eqref{eq:Lindblad} yields the Bloch equations for a mechanically driven CNT coupled to its environment (see Sec.~S2 in the Supplemental Material for details ~\cite{SM}): 
\begin{align}
\dot X&=-\Gamma_2 X+\Delta Y,\label{eq:BlochX}\\
\dot Y&=-\Gamma_2 Y-\Delta X+\Omega_R Z,\label{eq:BlochY}\\
\dot Z&=-\Gamma_1(Z-Z_{\!\mathrm{eq}})-\Omega_R Y,\label{eq:BlochZ}
\end{align}
with thermal inversion 
\(Z_{\!\mathrm{eq}} =(\gamma_{\uparrow}-\gamma_{\downarrow})/
    (\gamma_{\uparrow}+\gamma_{\downarrow}) = -\tanh\!\bigl(\hbar\omega_{0}/2k_{B}T\bigr)\) \cite{breuer_theory_2009}.
For an Ohmic spectral density at dilution temperatures one obtains $Z_{\!\mathrm{eq}}\approx-1$.

\textit{Physical picture --} The longitudinal relaxation time $T_1\equiv 1/\Gamma_1$ sets the population ringdown toward thermal equilibrium through energy-exchange processes with rates $\gamma_{\downarrow}$ and $\gamma_{\uparrow}$ (with $\Gamma_1=\gamma_{\downarrow}+\gamma_{\uparrow}$). The transverse coherence time $T_2\equiv 1/\Gamma_2$ governs the decay of the off-diagonal density-matrix elements in the Markovian limit and therefore sets the homogeneous contribution to the spectroscopic linewidth in the weak-drive limit (in contrast to additional inhomogeneous broadening from slow, quasi-static frequency fluctuations). In the relaxation-limited regime of negligible pure dephasing ($\gamma_{\varphi}\approx 0$), one has $\Gamma_2=\Gamma_1/2$ and hence $T_2\simeq 2T_1$ .

Eqs.~\eqref{eq:T1T2}--\eqref{eq:BlochZ} clarify the relevant control and decay scales: $\omega_{0}$ sets the intrinsic timescale, $\Omega_R$ controls the coherent rotation rate induced by AFM actuation, and the detuning $\Delta$ determines the Bloch-vector precession during free evolution (Ramsey phase accumulation). The transverse coherence time $T_2$ fixes the Ramsey-fringe contrast, while the longitudinal relaxation time $T_1$ governs the free decay of the population once the drive is switched off. Under continuous resonant drive, the driven steady state reflects competition between coherent control ($\Omega_R$) and decoherence ($\Gamma_1,\Gamma_2$), which sets the visibility of the driven response used below in \hyperref[sec:ramsey]{Section~IV} and \hyperref[sec:wigner]{Section~V}.

\begin{figure*}[ht]
    \centering
    \includegraphics[width=1\textwidth]{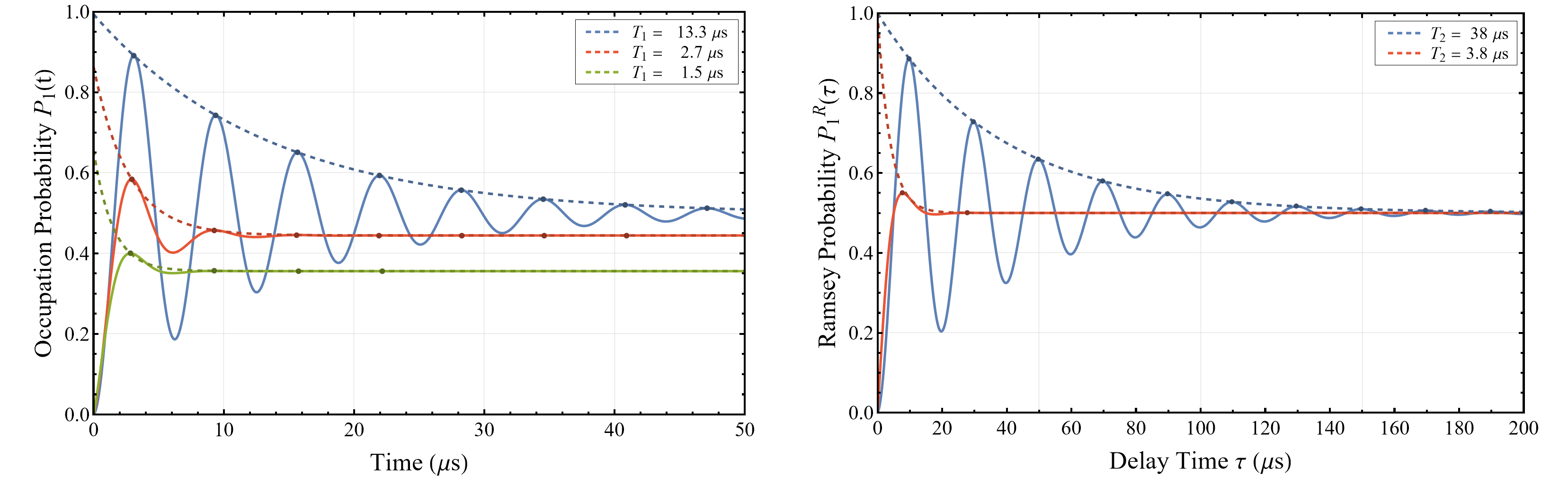}\\
    \vspace{0.3em}
    {\small (a) \hfill (b)}

\caption{(a) Excited-state population \(P_1(t)\) of a driven CNT in the effective two-level description for three representative relaxation strengths relative to the control rate:
  \(\Gamma_1/\Omega_R=0.1\) (weak relaxation; blue),
  \(\Gamma_1/\Omega_R=0.5\) (intermediate relaxation; red), and
  \(\Gamma_1/\Omega_R=0.9\) (strong relaxation; green).
  The Rabi frequency is \(\Omega_R=\SI{0.75}{MHz}\) for all curves.
  Dotted lines are fits to Eq.~\eqref{eq:P1} used to extract \(T_1\).
  (b) Ramsey population \(P_1^{\mathrm{R}}(\tau)\) versus delay time, for detuning \(\Delta=\SI{300}{kHz}\) in two CNT devices of different lengths:
  \(L=\SI{800}{\nano\meter}\) (blue) and \(L=\SI{250}{\nano\meter}\) (red).
  Dotted lines are fits to Eq.~\eqref{eq:PR} used to extract \(T_2\).}
\label{fig:probability_comparison}
\end{figure*}

\section{Rabi Oscillations and Ramsey Interferometry}\label{sec:ramsey}
On resonance ($\Delta=0$), at dilution temperatures, and with the system initialized in $\ket{0}$, the driven CNT qubit undergoes damped Rabi oscillations. Solving the Bloch equations ~\eqref{eq:BlochX}--\eqref{eq:BlochZ} and defining the damped Rabi frequency \(\widetilde\Omega =\sqrt{\Omega_R^2-(\Gamma_1-\Gamma_2)^2/4}\),  we obtain the following expression for the excited-state occupation probability (details in Sec.~S3 in the Supplemental Material \cite{SM}): 
\begin{equation}
\begin{aligned}
P_1(t)
&= \frac{\Omega_R^2}{2\big(\Omega_R^2+\Gamma_1\Gamma_2\big)}\cdot \\
&\cdot \Big[1-e^{-(\Gamma_1+\Gamma_2)t/2}\Big(\cos(\widetilde\Omega t)+\frac{\Gamma_1-\Gamma_2}{2\,\widetilde\Omega}\sin(\widetilde\Omega t)\Big)\Big],
\end{aligned}
\label{eq:P1}
\end{equation}
with the on-resonance steady state
\begin{equation}
P_1^{(\infty)}=\frac{\Omega_R^2}{2\big(\Omega_R^2+\Gamma_1\Gamma_2\big)}.
\end{equation}

In the relaxation-limited regime with negligible pure dephasing ($\gamma_{\varphi}\!\approx\!0$), one has $\Gamma_2\!\approx\!\Gamma_1/2$, and the Rabi oscillation envelope decays at rate $3\Gamma_1/4$. Additional pure dephasing increases $\Gamma_2$ while leaving $\Gamma_1$ unchanged, thereby reducing the oscillation visibility and shrinking the coherent oscillation window $\Omega_R>\tfrac{1}{2}|\Gamma_1-\Gamma_2|$ (for which $\widetilde\Omega$ remains real). The population ringdown time (measured after the drive is switched off) remains set by $T_1=1/\Gamma_1$. Fig.~\ref{fig:probability_comparison}(a) displays the time evolution of the excited-state occupation probability for different ratios $\Gamma_1/\Omega_R$, showing the crossover from resolved coherent oscillations ($\Gamma_1\ll\Omega_R$) to relaxation-dominated response ($\Gamma_1\lesssim\Omega_R$). The terminology refers to open two-level dynamics rather than to classical critical damping of a driven oscillator \cite{breuer_theory_2009}.

\textit{Ramsey interferometry---} This technique is a standard phase-sensitive protocol for quantifying coherence in an effective two-level system \cite{schlosshauer_decoherence_2007, breuer_theory_2009, wollack_quantum_2022}. Starting from an initial state near $\ket{0}$, a first $\pi/2$ pulse prepares a coherent superposition of $\ket{0}$ and $\ket{1}$ with a relative phase set by the drive. The system then undergoes free evolution for a variable delay time $\tau$, during which the relative phase accumulates at a rate determined by the detuning. A second $\pi/2$ pulse converts this accumulated phase into a measurable population difference, so that the $\ket{1}$ population oscillates as a function of $\tau$ (or, equivalently, as a function of a controlled phase shift of the second pulse). The resulting Ramsey fringes provide a direct diagnostic of coherence: the oscillation frequency and phase encode the detuning, while the decay of the fringe contrast yields the transverse coherence time. Ramsey measurements thus complement Rabi-type dynamics by providing a sensitive probe of dephasing and a route to extracting $T_2$ from the decay of the fringe contrast.

In our setup a calibrated AFM tip generates two \emph{mechanical} $\pi/2$ pulses separated by a free-evolution interval $\tau$. The pulse sequence is:
\[
    \ket{0}
    \xrightarrow{\pi/2}
    \frac{\ket{0}+\ket{1}}{\sqrt{2}}
    \xrightarrow{\text{free } \tau}
    \rho(\tau)
    \xrightarrow{\pi/2}
    \rho_\mathrm{out}(\tau)
    \xrightarrow{\text{readout}}
    P_1^{R}(\tau).
\]
where $\rho(\tau)$ denotes the reduced density matrix after the free-evolution interval (including decoherence), and $\rho_\mathrm{out}(\tau)$ is the density matrix after the second $\pi/2$ pulse. The first pulse implements a $\pi/2$ rotation via a short resonant drive of duration $t_{\pi/2}=\pi/(2\Omega_R)$ and prepares the CNT in the superposition $\ket{\psi(0)}=(\ket0+\ket1)/\sqrt2$. During the subsequent delay interval $\tau$ the Bloch vector precesses at the detuning frequency $\Delta$, and the system acquires a relative phase $\Delta \tau$ induced by the unitary evolution. The second $\pi/2$ pulse converts the accumulated phase into a population difference, yielding the Ramsey excited-state probability (detailed derivation provided in Sec.~S3 in the Supplemental Material \cite{SM}): 
\begin{equation}
P_1^{\mathrm{R}}(\tau)=\tfrac{1}{2}\Big[1+e^{-\tau/T_2}\cos(\Delta\tau+\phi_0)\Big],
\label{eq:PR}
\end{equation}
where $\phi_0$ captures any programmed phase shift between the two AFM pulses. In our protocol $P_1^{\mathrm{R}}(\tau)$ is measured immediately after the second pulse using any of the compatible transducers  described in \hyperref[sec:wigner]{Section~V}  (AFM deflection, dispersive Cooper-pair box, or cavity reflectometry). Sweeping $\tau$ produces Ramsey fringes whose frequency is set by the detuning $\Delta$ and whose envelope decays with $T_2$. We operate with $\Omega_R, \Delta \ll \omega_{0}$ (RWA regime) and use short pulses, $t_{\pi/2}\ll \tau, T_2$, so decoherence during the pulses is negligible. The relative phase $\phi_0$ can be set either by timing the second pulse with respect to the drive or by an explicit phase shift of the control waveform. On resonance ($\Delta=0$) this enables phase scans at fixed $\tau$. Population measurement in the \(\left\{\ket{0},\ket{1}\right\}\) manifold then returns  $P_1^{\mathrm{R}}(\tau)$ with explicit dependence on the accumulated phase \(\Delta \tau\) and the controlled pulse phase $\phi_0$ (Eq.~\eqref{eq:PR}). Fig. \ref{fig:probability_comparison}(b) shows the Ramsey probability as a function of the delay time for representative device parameters. The $\tau$-dependent modulation of the measurement statistics provides direct evidence of a coherent superposition of the two levels. The energy-relaxation time $T_1$ can be obtained independently from population ringdown, after switching off the drive, with $T_1=Q/\omega_{0}$, for a mode of quality factor $Q$ \cite{schneider_observation_2014}. This all-mechanical implementation eliminates on-chip microwave drive lines at the resonator, avoids optical heating, and uses the same actuator to deliver phase-space displacements for Wigner tomography, thereby unifying control and state reconstruction in a single device, as we show in \hyperref[sec:wigner]{Section~V}.

\begin{figure*}[t]
    \centering
    \begin{minipage}{0.88\textwidth}
        \includegraphics[width=0.28\textwidth]{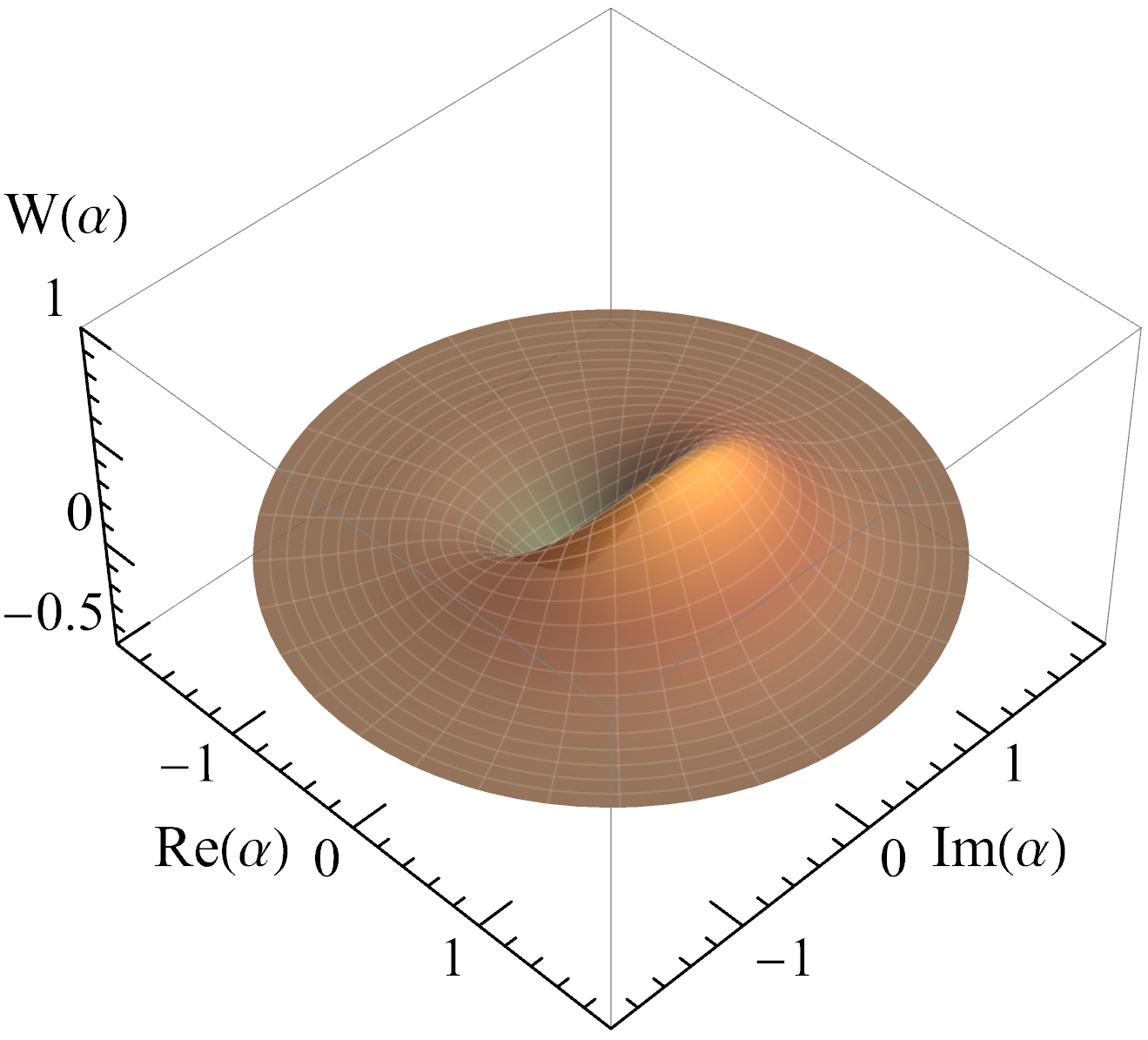}\hfill
        \includegraphics[width=0.28\textwidth]{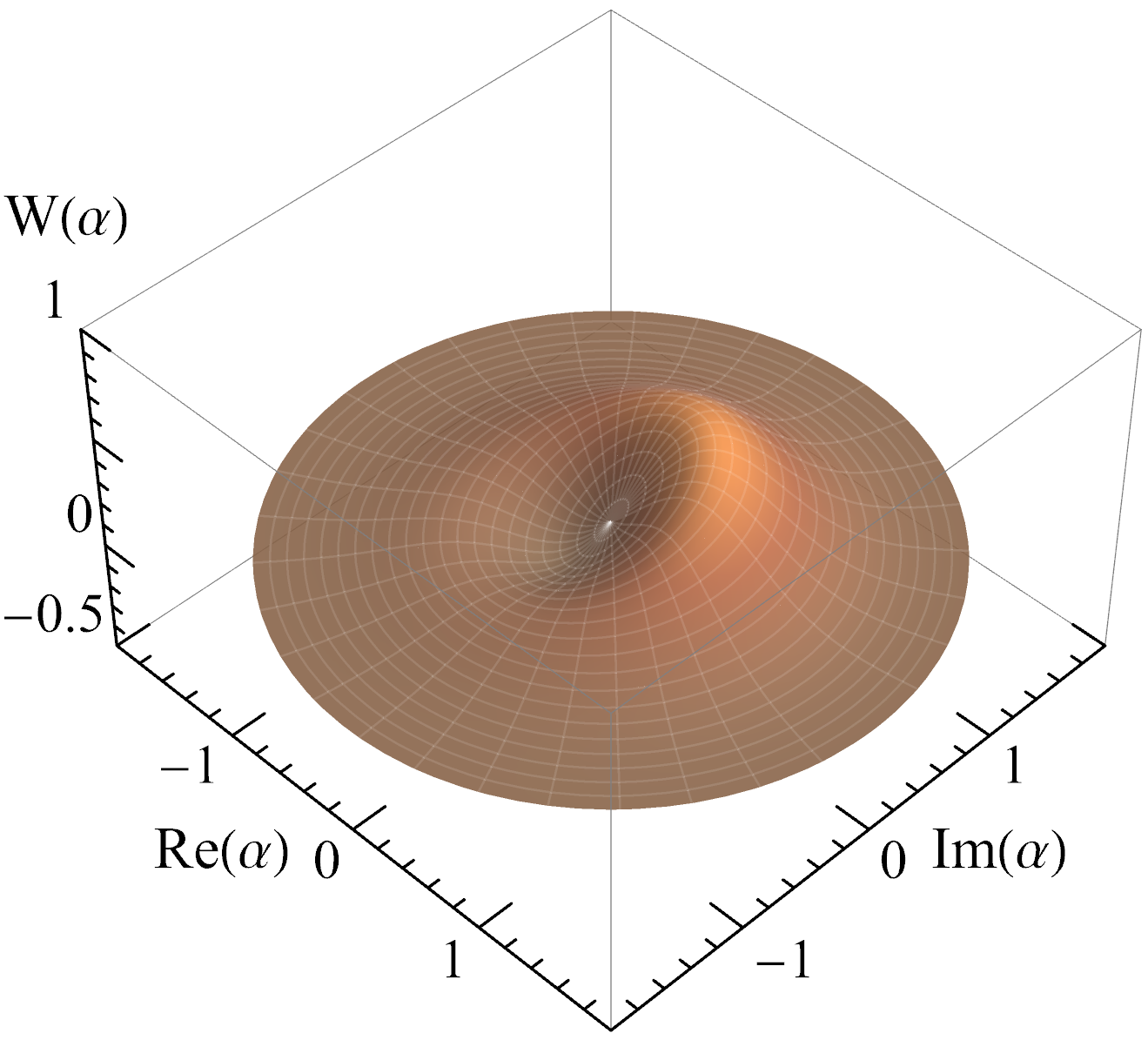}\hfill
        \includegraphics[width=0.28\textwidth]{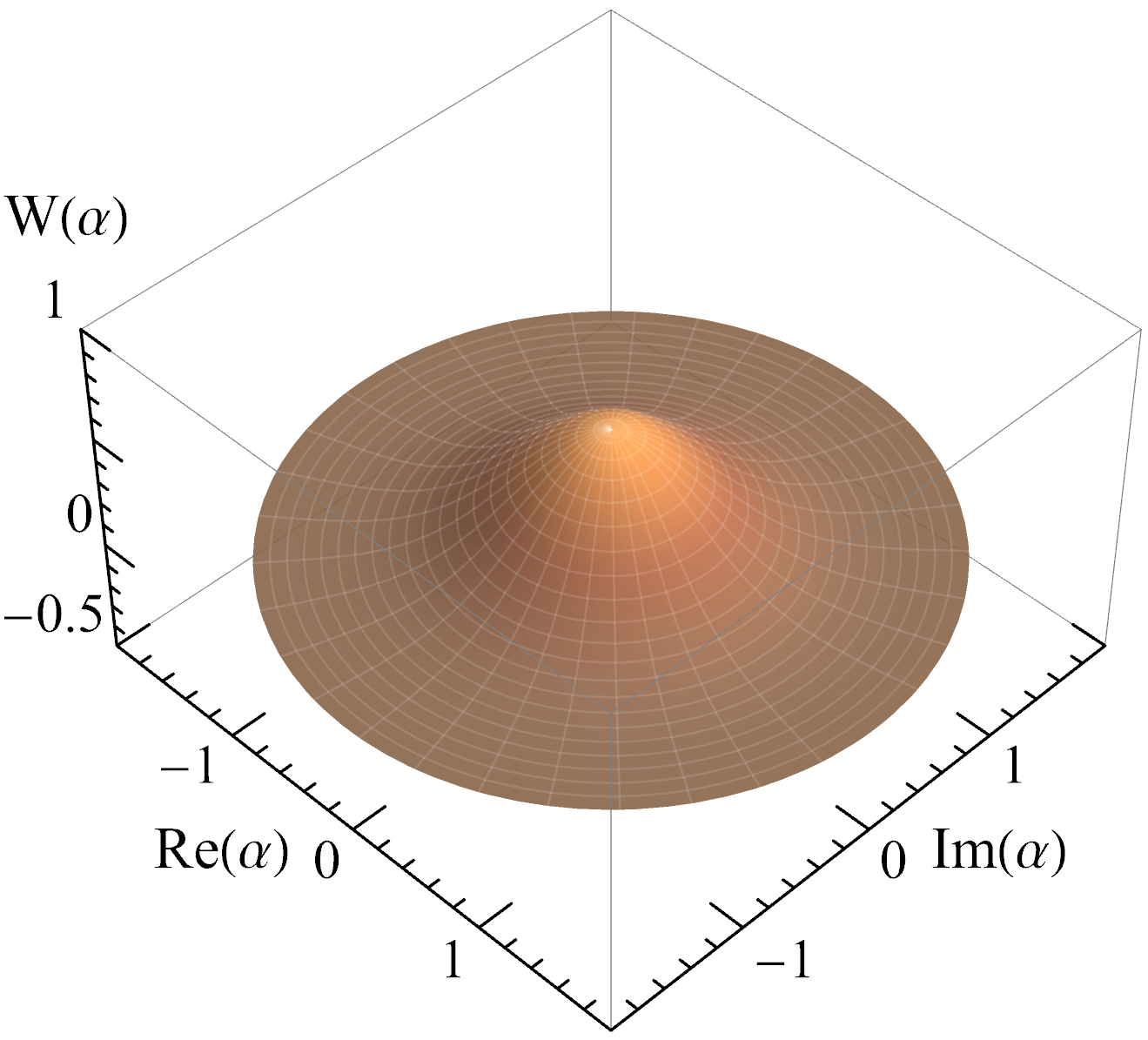}\\
        \vspace{0.3em}
        {\small (a) \hspace{0.3em} \hfill(b)\hfill(c)}

        \vspace{0.9em}

        \includegraphics[width=0.28\textwidth]{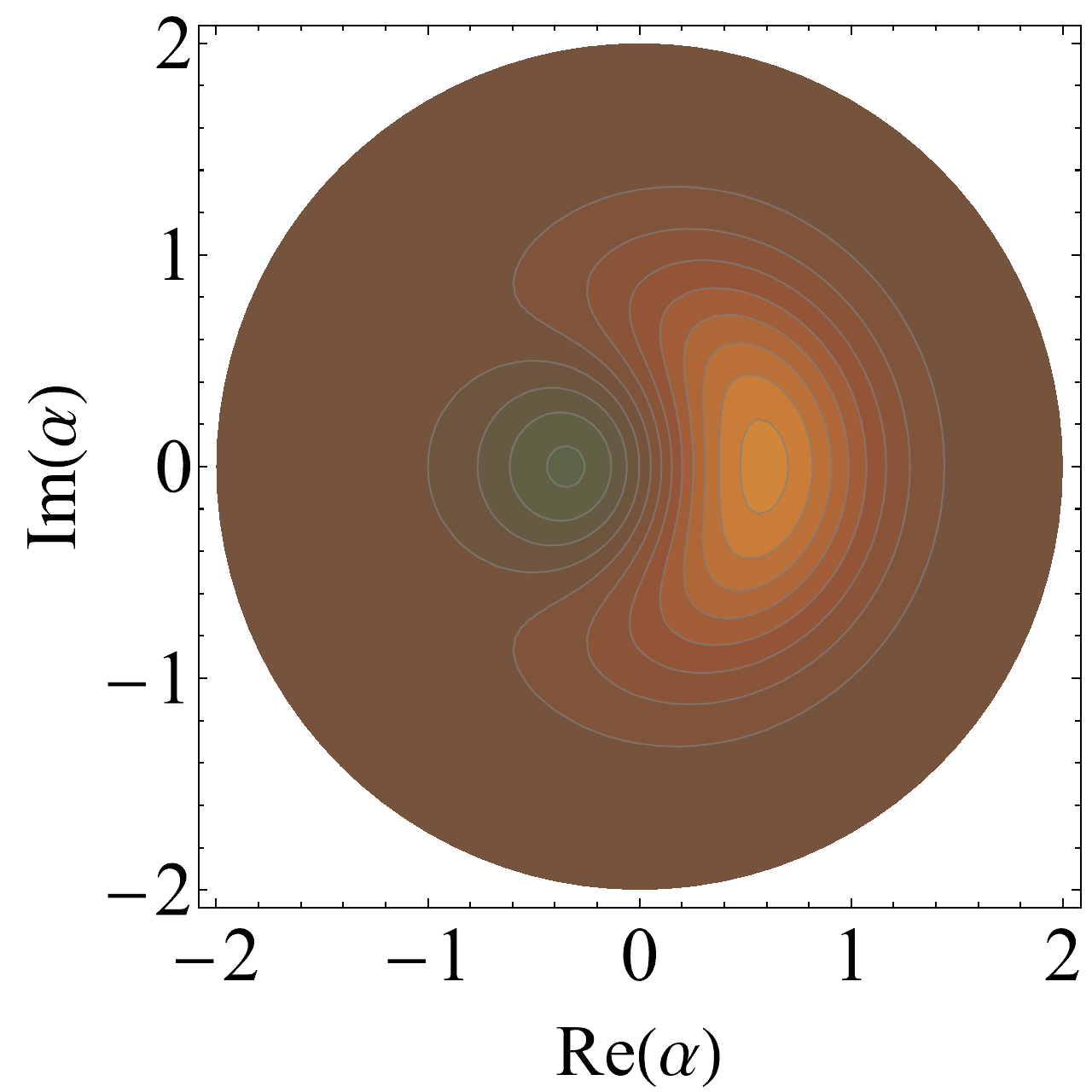}\hfill
        \includegraphics[width=0.28\textwidth]{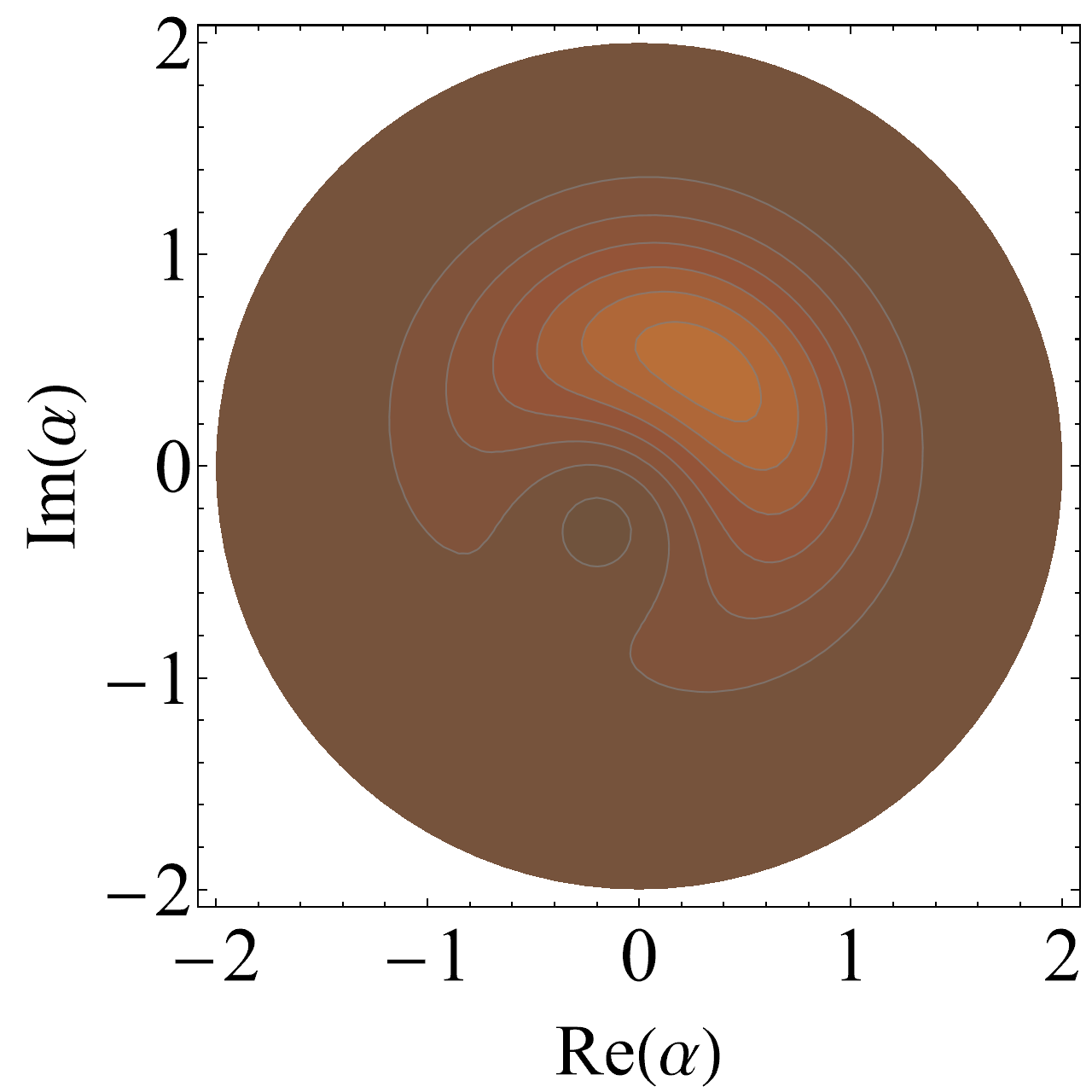}\hfill
        \includegraphics[width=0.28\textwidth]{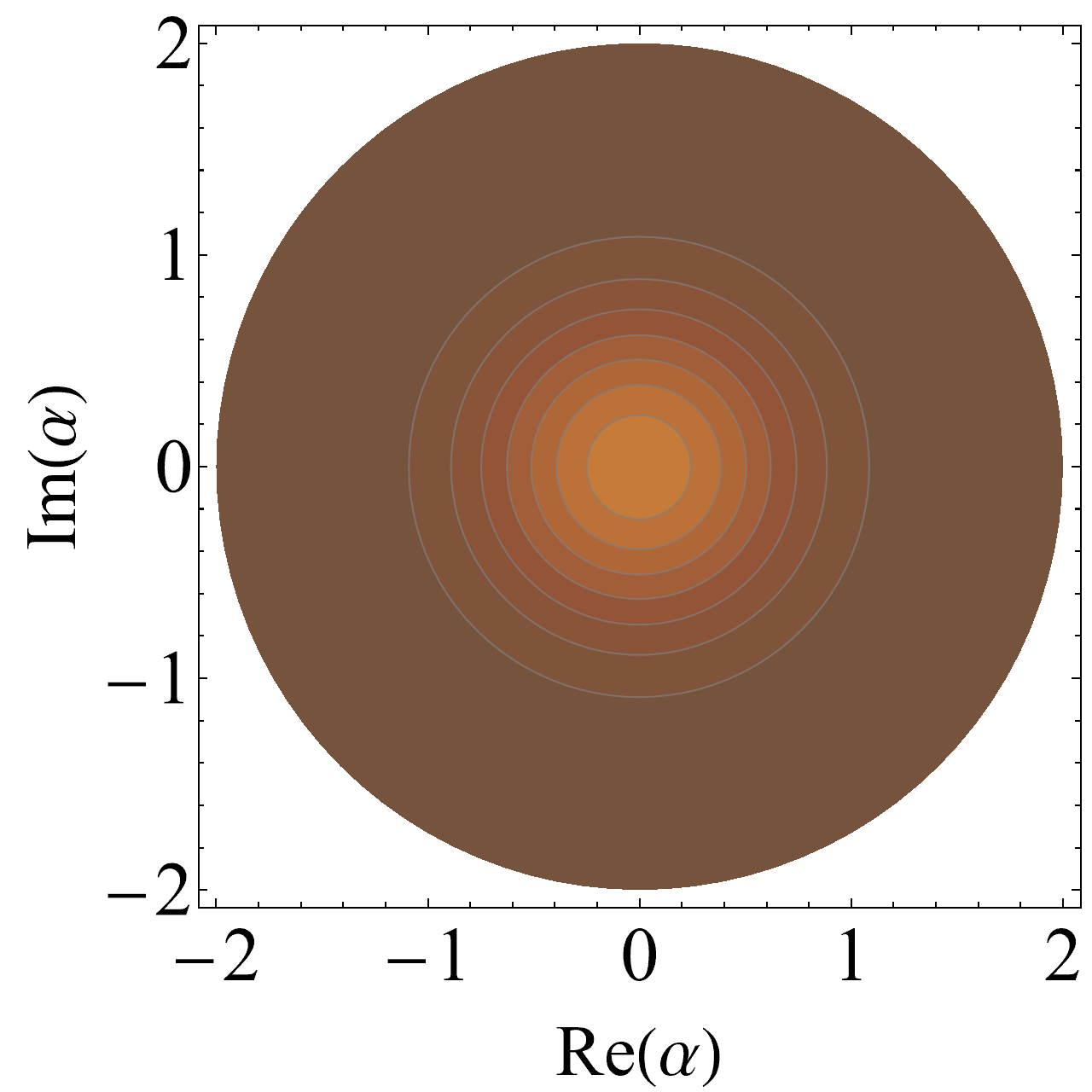}\\
        \vspace{0.3em}
        {\small (d) \hspace{1.2em} \hfill(e)\hfill(f)}
    \end{minipage}\hfill
    \begin{minipage}{0.12\textwidth}
        \centering
        \includegraphics[height=0.3\textheight]{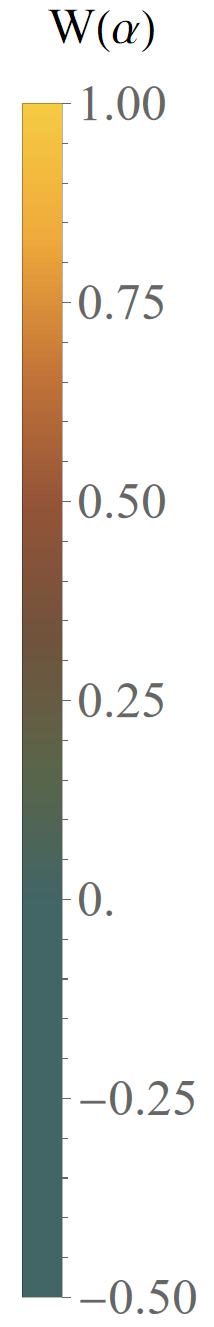}
    \end{minipage}

    \caption{Time evolution of the calculated Wigner function \(W(\alpha)\).
    For all panels, \(T_1 = 5/\omega_{0}\). Top row: 3D surfaces showing \(W(\alpha)\) at \(t=0\) (a), \(t=T_1\) (b), and \(t=5T_1\) (c). Bottom row: corresponding 2D plots at \(t=0\) (d), \(t=T_1\) (e), and \(t=5T_1\) (f). A single color scale is shown on the right, applying to all six panels.}
    \label{fig:decoherence_comparison}
\end{figure*}

\section{Wigner Tomography}\label{sec:wigner}

The Wigner function $W(\alpha)$ provides a complete quasi-probability representation of a bosonic mode in phase space, carrying the same information as the density matrix \cite{weinbub_recent_2018,case_wigner_2008}. Unlike a classical probability distribution, $W(\alpha)$ can assume negative values, a widely used and experimentally relevant signature of nonclassicality associated with quantum interference. Thus the Wigner function is a particularly intuitive tool for visualizing state preparation and decoherence. Experimentally, reconstructing $W(\alpha)$ (``Wigner tomography'') requires repeated measurements on identically prepared states. One common route is to measure a family of rotated quadrature distributions (marginals) and reconstruct $W(\alpha)$ by inversion procedures such as an inverse Radon transform \cite{weinbub_recent_2018, case_wigner_2008}.
 An equivalent and often more direct route is displaced-parity tomography: one first applies a controlled phase-space displacement by an amount $\alpha$, and then measures the parity of the displaced state \cite{bertet_direct_2002,leibfried_experimental_1996}. 
 Because the Wigner function is proportional to the expectation value of the displaced parity operator, scanning $\alpha$ across phase space yields $W(\alpha)$ point-by-point, without requiring a global inversion. These approaches are well established in platforms such as matter-wave interferometry and trapped-ion motion \cite{wang_creating_2016, wollack_quantum_2022, lougovski_fresnel_2003, bertet_direct_2002, leibfried_experimental_1996, karlovets_possibility_2017, weinbub_recent_2018}.

In our protocol the AFM tip supplies the required controlled displacements of the CNT mode, while parity information is first mapped (via the dispersive schemes described below) onto a two-level system population measurement, enabling reconstruction of $W(\alpha)$ and direct identification of nonclassical features, including regions where the Wigner function is negative.

\textit{AFM-implemented displacements---} A short phase-controlled AFM impulse implements the phase-space displacement
\begin{equation}
D(\alpha)=\exp(\alpha a^\dagger-\alpha^* a),\qquad \alpha=\frac{F_0 t_d}{\sqrt{2 m_{\!\mathrm{eff}}\hbar\omega_{0}}}\,e^{i\phi_d},
\label{eq:Dalpha}
\end{equation}
where \(\alpha=(X+iP)/\sqrt2\) is the phase-space coordinate, \(F_0\) and \(t_d\) are the pulse amplitude and duration, \(\phi_d\) is the phase of the AFM drive relative to the oscillator, and \(m_{\!\mathrm{eff}},\, \omega_0\) are the effective mass and frequency of the CNT fundamental flexural mode. In practice, $\alpha$ is calibrated experimentally from the measured driven response (or equivalently from the known drive waveform together with $m_{\!\mathrm{eff}}$ and $\omega_0$).  After the calibrated displacement, measuring the parity operator $\hat{\Pi}=e^{i\pi a^\dagger a}=(-1)^{\hat n}$ yields the displaced-parity form of the Wigner function \cite{bertet_direct_2002, lougovski_fresnel_2003, leibfried_experimental_1996}:
\begin{equation}
W(\alpha)= \frac{2}{\pi}\langle\hat\Pi\rangle_{\alpha}  =\frac{2}{\pi} \mathrm{Tr}\big[D(-\alpha)\,\rho_S\,D(\alpha)\,\hat\Pi\big]\quad
\label{eq:Wpar}
\end{equation}

\textit{Parity mapping and measured signal---} The same AFM actuator that generates \(\pi/2\) rotations (Rabi/Ramsey calibration) also provides the controlled phase-space displacements $D(\alpha)$ required for Wigner tomography. Parity is converted to a measured excited state population by a Ramsey-type sequence \(\{\pi/2,\ t_\pi,\ -\pi/2\}\) that accumulates a number-dependent phase $\chi \hat{n}t_{\pi}$. Here, $-\pi/2$ denotes a $\pi/2$ pulse with a $\pi$-phase flip of the drive. Choosing \(t_\pi=\pi/\chi\) implements the parity phase gate \(e^{i\pi\hat n}\). The parameter $\chi$ is the effective parity-mapping rate per phonon set by the CPB-CNT or cavity-CNT interactions (details below and in Sec.~S6 the Supplemental Material \cite{SM}). This mapping yields the excited--state probability of the TLS \cite{bertet_direct_2002, lougovski_fresnel_2003, leibfried_experimental_1996}:
\begin{equation}
P_1(\alpha)=\tfrac12\big[1- \frac{\pi}{2}W(\alpha)\big],\qquad 2\pi/\omega_{0}\ll t_\pi\ll T_2.
\label{eq:ProbW}
\end{equation}

Sweeping \(\alpha\) on a grid maps \(W(\alpha)\) directly via displaced-parity measurements, without any inverse transform or numerical reconstruction. The contrast across the map is limited by \(T_2\) during the parity mapping window \(t_\pi\) and by any displacement-induced dephasing.

Fig.~\ref{fig:decoherence_comparison} shows the time evolution of the calculated $W(\alpha)$ for the CNT (details in Sec.~S5 in the Supplemental Material \cite{SM}). Regions with $W(\alpha)<0$ are a hallmark of nonclassicality arising from interference between coherently superposed vibrational states and are clearly visible at early times (panels (a), (b), (d), (e)). As the CNT mode evolves under environmental decoherence (energy relaxation and dephasing), these interference features are progressively washed out, and $W(\alpha)$ approaches a positive, broadened Wigner distribution that tends to the ground state Gaussian at long times (panels (c) and (f)).

\textit{Readout modalities---}
We consider four complementary readout schemes: (a) AFM-based deflection (direct mechanical transduction), (b) dispersive Cooper-pair-box (CPB) readout with parity mapping, (c) dispersive microwave-cavity readout of the CPB, and (d) direct cavity readout of the CNT mechanics, without an intermediate CPB. All four methods share the same mechanical basic controls (AFM-driven $\pi/2$ rotations and calibrated phase-space displacements), extending the Ramsey interferometer into a full, state-resolved characterization of the phase space. We summarize these modalities below
and give full details of their implementation, including
calibration, timing, and acquisition in Secs.~S6-S7 of the Supplemental Material \cite{SM}. Parity-resolved Wigner tomography in our protocol relies on the dispersive parity-mapping schemes (b)--(d) below, whereas AFM deflection in (a) provides direct mechanical transduction primarily for spectroscopy and Rabi/Ramsey calibration.

\emph{(a) AFM deflection and direct transduction---}
After the second pulse in the Ramsey control sequence, the CNT displacement is transduced by the AFM probe (via optical/electrical deflection readout or a weak probe force) through its coupling to the CNT displacement operator $\hat x=x_{\rm zpf}(a+a^\dagger)$ with the Hamiltonian $H_{\rm meas}(t)=-F_{\rm probe}(t)\,\hat x$. Demodulation of the probe response at $\omega_d\simeq\omega_{0}$ provides access to the mechanical quadrature amplitude/phase.
In this direct-transduction channel the demodulated AFM signal measures a mechanical quadrature (equivalently the driven response amplitude/phase), which is converted to an inferred ensemble-averaged $P_1$ via calibration against Rabi/Ramsey scans in the effective two-level operating regime (see Sec.~S6 in the Supplemental Material~\cite{SM}). In the same channel, ringdown of the demodulated response determines $T_1$, while sweeping the Ramsey delay time $\tau$ yields the detuning frequency $\Delta$ and the coherence time $T_2$ (Eq.~\eqref{eq:PR}).

\emph{(b) Dispersive CPB readout and parity mapping---}
A CPB dispersively coupled to the CNT implements a phase proportional to the phonon-number operator $\hat n=a^\dagger a$ during the calibrated Ramsey free-evolution window ~\cite{armour_entanglement_2002, blencowe_quantum_2004}.
In the dispersive regime, the effective shift per phonon $\chi$ enables a parity gate $e^{i\pi\hat n}$ by choosing $t_\pi=\pi/\chi$.
Preceded by a controlled displacement $D(\alpha)$ and combined with the pulse sequence $\{\pi/2,\,t_\pi,\,-\pi/2\}$, the final CPB population encodes the displaced-parity signal and thus $P_1(\alpha)$ (and equivalently $W(\alpha)$) (details in Secs. ~S6 and S7 in the Supplemental Material ~\cite{SM}).
The displacement phase is set by the relative phase between the displacement waveform and the mechanical reference at $\omega_d$, and the contrast is limited primarily by decoherence during $t_\pi$. For typical CNT resonators \((L\sim 0.1\text{–}1~\mu\mathrm{m})\), this corresponds to \(t_{\pi}\sim 1\text{–}20~\mu\mathrm{s}\) (see \hyperref[sec:estimates]{Section~VI}).
Mechanical implementations of such dispersive electromechanical couplings have been demonstrated with CPB/transmon devices~\cite{lahaye_nanomechanical_2009, pirkkalainen_cavity_2015, lee_strong_2023}.

\emph{(c) Dispersive cavity readout of the CPB---}
In this scheme an additional microwave cavity operated in the dispersive regime provides phase-sensitive readout of the CPB qubit state (and thus of the Ramsey/parity-mapped outcome) through a qubit--cavity dispersive pull, as in circuit QED~\cite{regal_measuring_2008, teufel_circuit_2011}.
Standard homodyne detection of the transmitted/reflected cavity field resolves \(\ket{0}\) vs \(\ket{1}\) and converts the mapped CPB observable into a voltage record  Combined with the parity mapping this returns \(W(\alpha)\) (see Sec.~S6 in the Supplemental Material~\cite{SM}).

\emph{(d) Direct cavity--mechanics readout---}
As an alternative to CPB-based readout, the cavity can couple directly to the CNT mode to produce either a cross-Kerr \cite{aspelmeyer_cavity_2014}, or a radiation-pressure interaction \cite{blais_circuit_2021, aspelmeyer_cavity_2014}. During the Ramsey free-evolution window, these couplings generate an effective phase rate $\chi_{\rm eff}$ proportional to the CNT phonon number $\hat n$, so the same parity gate $e^{i\pi\hat n}$ is realized by choosing $t_\pi=\pi/\chi_{\rm eff}$ \cite{SM}. The tomography framework and normalization are the same as described above.

\section{Estimated device parameters and AFM-CNT interactions} \label{sec:estimates}
\textit{Geometry, material constants, and assumptions --} We model the suspended CNT as a uniform Euler–Bernoulli beam of length \(L\), radius \(r\), cross-section area $A$ and moment of inertia $I$. Material parameters used throughout are Young’s modulus \(E=1\)TPa and mass density \(\rho=2200\ {\text{kg}/\text{m}^{-3}}\).
For a thin-walled nanotube \(I/A\!\simeq\!r^2/2\), yielding a fundamental flexural frequency \cite{garcia-sanchez_mechanical_2007}:
\begin{equation}
\omega_0 \simeq \frac{\beta_1^2 r}{\sqrt{2}\,L^2}\sqrt{\frac{E}{\rho}}.\label{eq:w0tube}
\end{equation}
with \(\beta_{1}\approx 4.73\) for the fundamental mode. The linear frequency is \(f_{0}=\omega_{0}/2\pi\) and scales as \(f_{0}\propto L^{-2}\).

\textit{Effective mass and zero-point motion -- } Let \(\phi_{1}(x)\) be the normalized fundamental mode shape and choose the readout/drive coordinate as the midpoint displacement \(q(t)=w(x=L/2,t)\).
The associated effective mass is
\begin{equation}
m_{\mathrm{eff}}=\mu\int_{0}^{L}\!\frac{\phi_{1}(x)^{2}}{\phi_{1}(L/2)^{2}}\,dx
\equiv c_{\mathrm{c}}\,\rho A L,
\label{S:meff_def}
\end{equation}
where \(c_{\mathrm{c}}\simeq 0.735\) is a dimensionless geometric factor for the double-clamped beam \cite{garcia-sanchez_mechanical_2007}. The zero-point motion of this mode is then
\begin{equation}
x_{\mathrm{zpf}}=\sqrt{\frac{\hbar}{2\,m_{\mathrm{eff}}\,\omega_{0}}}
= \sqrt{\frac{\hbar}{2\,c_{\mathrm{c}}\,\rho A L\,\omega_{0}}}\;\propto\; L^{1/2}.
\label{S:xzpf}
\end{equation}

As discussed in the previous sections, an AFM force $F(t)$ couples as $-F(t) \,\hat x$, giving a Rabi frequency \(\Omega_R(t)= F(t)\,x_{\mathrm{zpf}}/\hbar\). In the weakly anharmonic regime relevant here $x_{01}\equiv\langle0|\hat x|1\rangle\simeq x_{\rm zpf}$, and the TLS Rabi rate may be estimated using $x_{\rm zpf}$. 
 To realize a \(\pi/2\) pulse of duration \(t_{\pi/2}\) one requires $\Omega_R = \pi/({2 t_{\pi/2}})$ and therefore an applied force $F_{\pi/2} \simeq \pi \hbar /(2t_{\pi/2}\cdot x_{\mathrm{zpf}})$. Thus, the force scale is set by \(x_{\mathrm{zpf}}\) (decreasing as \(\sqrt L\) for fixed geometry) and is conveniently calibrated from a Rabi scan. Ramsey fringes yield the detuning \(\Delta\) and transverse coherence time \(T_2\). Ringdown of the demodulated AFM signal returns \(T_1=Q/\omega_0\). Taking \(Q=10^4\) (reported for suspended CNTs at tens of mK \cite{Steele2009Science, CirioPhysRevLett.109.147206}), $r=1$ nm, and \(t_{\pi/2}=0.1\) $\mu$s we calculate the parameter scalings with CNT length \(L\) as summarized in Table~\ref{tab:parameters} below:
\begin{table}[h]
    \centering
    \small
    \begin{tabular}{@{}c@{\hspace{24pt}}c@{\hspace{24pt}}c@{\hspace{24pt}}c@{\hspace{24pt}}c@{\hspace{24pt}}c@{}}
        \hline\hline
        \(L\) & \(\omega_0/2\pi\) & \(x_{\mathrm{zpf}}\) & \(F_{\pi/2}\) & \(T_1\)  & \(T_2\)  \\
        (nm)  & (MHz)        & (pm)                 & (fN)        & ($\mu$s) & ($\mu$s) \\
        \hline
        100   & 5370        & 2.14                 & 0.77        & 0.29     & 0.58     \\
        500   & 221        & 4.79                 & 0.35        & 7.4     & 14.8      \\
        1000  & 54       & 6.78                 & 0.24        & 29.6      & 59.2      \\
        \hline\hline
    \end{tabular}

    \caption{Representative parameters as a function of CNT length. Calculated values use \(Q=10^4\), \(t_{\pi/2}=\SI{0.1}{\mu}s\) and the Euler--Bernoulli beam model discussed in the main text. Here \(T_2\simeq 2T_1\) assuming negligible pure dephasing.
} 
    \label{tab:parameters}
\end{table}

Because the force required for $\pi/2$ pulses scales as $F_{\pi/2}\propto L^{-1/2}$, mechanically driven rotations require a weaker drive in longer tubes. For fixed $Q$ and negligible pure dephasing, the energy-relaxation time grows as \(T_1=Q/\omega_0\propto L^2\). The trade-off is a lower \(\omega_0\) (hence higher thermal occupation at a given temperature) for larger \(L\). Short CNTs favor ground-state occupation at \SI{10}{mK}, whereas longer CNTs favor extended coherence. The table quantifies this balance.

As a simple consistency check on Eq.~\eqref{eq:w0tube}, one may combine the well-known static midpoint stiffness of a fixed--fixed Euler--Bernoulli beam under a point load,
$k_{\mathrm{mid}} = 192\,E I/L^{3}$, with the effective mass for the midpoint coordinate $m_{\mathrm{eff}}=c_{c}\rho A L$ used above. This yields an estimate
$\omega_{\mathrm{est}}\simeq \sqrt{k_{\mathrm{mid}}/m_{\mathrm{eff}}}
= \sqrt{192/c_c}\,L^{-2}\sqrt{E I/(\rho A)}$, which has the same $L^{-2}$ scaling as Eq.~\eqref{eq:w0tube} but a slightly smaller numerical prefactor. Consequently, the stiffness-based estimate gives $\omega_{\mathrm{est}}\approx 0.72\, \omega_{0}$ for our parameters, as expected since $k_{\mathrm{mid}}$ characterizes a static point-load compliance rather than the full distributed eigenmode problem. 

\textit{Magnitude and scaling of the anharmonicity.---}
Intrinsic geometric nonlinearity in a doubly clamped CNT produces an effective quartic correction to the elastic restoring potential. In the quantum description this yields a Kerr-type self-anharmonicity \(K\) that can be strongly enhanced by electrostatic softening \cite{rips_hartmann_prl_2013, zueco_qubit_2009}. A concrete benchmark is provided by Ref.~\cite{rips_hartmann_prl_2013}, which reports that softening a CNT mode to $\omega_0/2\pi\simeq\SI{26.6}{MHz}$ yields an effective anharmonic separation $(\omega_{21}-\omega_{10})/2\pi\simeq\SI{2.71}{MHz}$, i.e.\ an effective MHz-scale anharmonicity in a realistic nanotube geometry. In that regime, the two-level selectivity conditions are readily satisfied: \( |K|\gg \Gamma_2\) for \(Q\sim 10^4\), since $\Gamma_2\simeq\Gamma_1/2= \omega_0/(2Q)\sim 10-100~\mathrm{kHz}$, and \( |K|\gg \Omega_R\) is ensured by choosing control bandwidths below the anharmonic separation (or equivalently, using longer and spectrally narrower pulses).

\textit{Range of validity for the TLS approximation.---}
The effective two-level description employed throughout the manuscript assumes an operating point where (i) the mechanical spectrum is sufficiently anharmonic that a near-resonant drive addresses the $|0\rangle\!\to\!|1\rangle$ transition without appreciable leakage to higher levels, and (ii) the mode is initialized close to its ground state. As discussed above, condition (i) is ensured by working with an anharmonic mechanical Hamiltonian and by enforcing the spectral selectivity criterion in Eq.~\eqref{eq:tls_conditions}, which is standard in nanomechanical-qubit proposals \cite{rips_hartmann_prl_2013,wang_method_2016,sarma_tunable_2018}. Condition (ii) requires either $\hbar\omega_{0}\gg k_B T$ (for sufficiently high frequency) or active cooling to an effective phonon occupancy $\bar n_{\mathrm{th}}\ll 1$ (for lower-frequency devices), as is routinely assumed in quantum nanomechanics \cite{schwab_putting_2005,lahaye_approaching_2004,WilsonRae2007PRL,Urgell2020NatPhys}.
For a mode of frequency $\omega_0$ in equilibrium at temperature $T$, the thermal occupation is given by the Bose factor $\bar n_{\mathrm{th}}(T)=\bigl[e^{\hbar\omega_0/k_BT}-1\bigr]^{-1}$.
At \(T\sim 10~\mathrm{mK}\), GHz-frequency CNT modes are near the ground state, while MHz-scale modes have thermal occupation and therefore require sideband cooling to reach \(\bar n_{\mathrm{th}}\ll 1\). For example, using the representative frequencies in Table~\ref{tab:parameters}, one finds \(\bar n_{\mathrm{th}}\simeq 6\times 10^{-12}\) for \(\omega_0/2\pi=5.37~\mathrm{GHz}\), and \(\bar n_{\mathrm{th}}\simeq 0.5\) (or \(\simeq 3.4\)) for \(\omega_0/2\pi=221~\mathrm{MHz}\) (\(54~\mathrm{MHz}\)), respectively, thus requiring sideband cooling \cite{teufel_sideband_2011, oconnell_quantum_2010}. In summary, by combining the above we conclude that the effective TLS model is quantitatively controlled in a parameter window where (1) the mode is prepared with $\bar n_{\mathrm{th}}\ll 1$ (either intrinsically at high $\omega_0$ or via active cooling), and (2) the anharmonic separation exceeds both the total linewidth and the characteristic control bandwidth (Eq.~\eqref{eq:tls_conditions}). Derivation of the $K$ estimates and their scaling, together with strategies for active cooling, numerical benchmarks and a leakage estimate in terms of $\Omega_R/|K|$, is provided in Sec.~S9 in the Supplemental Material \cite{SM}.

\textit{AFM--CNT interactions --}
The AFM tip interacts with a suspended CNT through a combination of electrostatic and dispersion (van der Waals) forces. In practical AFM operation, these interactions can span a wide range: contact mode forces are typically in the nN regime, while noncontact/dynamic force microscopy operation is specifically designed to minimize tip--sample forces and dissipation \cite{giessibl_advances_2003}. Our protocol is intended to operate in ultrahigh vacuum and in \emph{noncontact} regime in which (i) the tip--CNT separation is held at a controlled working distance $z_0$ (tens of nm), (ii) the static interaction force is compensated (or simply treated as a constant offset that shifts the equilibrium position), and (iii) only a \emph{small}, calibrated \emph{AC} force component near the CNT mechanical frequency is used for coherent control.
The AFM can influence the CNT quantum dynamics primarily through small fluctuations of the tip--CNT working distance, $\delta z(t)$, and of the tip bias. The separation distance noise modulates the force gradient and thus the CNT mode frequency, contributing to dephasing. Bias noise can generate force noise near resonance and thus drive unwanted excitation. Importantly, in our scheme the resonant drive is supplied by \emph{electrical} modulation at $\omega_d\simeq\omega_0$, so the required control forces can be in the fN and sub-fN range even in the presence of a much larger static background interaction.

Table~\ref{tab:parameters} shows that coherent $\pi/2$ rotations require sub-fN force amplitudes. This is feasible because the relevant quantity is the \emph{AC} force component at $\omega_d\simeq\omega_0$, which can be made small and precisely tunable by modulating the tip bias (or a nearby gate) rather than mechanically oscillating the cantilever \cite{garcia-sanchez_mechanical_2007}. A minimal electrostatic model is \cite{garcia-sanchez_mechanical_2007, san-paulo_detection_2007, garcia_emergence_2012}
\begin{equation}
F_{\mathrm{el}}(t)\simeq \frac{1}{2}\frac{dC(z_0)}{dz}\,V(t)^2,
\qquad
V(t)=V_{\mathrm{dc}}+V_{\mathrm{ac}}\cos(\omega_d t),
\label{eq:Fel}
\end{equation}
which yields an AC force component near $\omega_d$ of amplitude
\begin{equation}
F_{\omega_d}\;\simeq\;\left|\frac{dC}{dz}\right|\,V_{\mathrm{dc}}V_{\mathrm{ac}}.
\label{eq:Fomega}
\end{equation}
Here $V_{\mathrm{dc}}$ is the static (dc) tip--CNT bias at the working point, and $V_{\mathrm{ac}}$ is the amplitude of an rf modulation applied at angular frequency $\omega_d$ (chosen $\omega_d\simeq\omega_0$ for resonant actuation). Equation~\eqref{eq:Fomega} assumes $V_{\mathrm{ac}}\ll V_{\mathrm{dc}}$, so the response at $\omega_d$ is dominated by the cross term $\propto V_{\mathrm{dc}}V_{\mathrm{ac}}$ (the $V_{\mathrm{ac}}^2$ term contributes primarily at $2\omega_d$ and as a small dc shift \cite{garcia-sanchez_mechanical_2007, san-paulo_detection_2007, garcia_emergence_2012}).

For an order-of-magnitude estimate, taking an effective capacitive area $A_{\mathrm{eff}}\sim(50~\mathrm{nm})^2$ and $z_0\sim 50~\mathrm{nm}$, gives
$|dC/dz|\sim \varepsilon_0 A_{\mathrm{eff}}/z_0^2 \sim 10^{-11}~\mathrm{F/m}$.
Voltages readily achievable in cryogenic rf setups, e.g.\ $V_{\mathrm{dc}}\sim 0.1~\mathrm{V}$ and $V_{\mathrm{ac}}\sim 1~\mathrm{mV}$, then produce
$F_{\omega_d}\sim |dC/dz|\,V_{\mathrm{dc}}V_{\mathrm{ac}}\sim (10^{-11}\times 10^{-4})~\mathrm{N}\sim 1~\mathrm{fN}$,
i.e.\ directly in the fN (and, with slightly smaller voltages, sub-fN) range required for coherent control, while remaining far below typical contact-mode forces. We emphasize that this separation of quasi-static positioning from high-frequency electrical actuation has already been implemented in AFM-based measurements of suspended CNT/graphene resonators \cite{garcia-sanchez_mechanical_2007, garcia-sanchez_imaging_2008}, and multifrequency dynamic-force microscopy enables vibration detection on rf devices up to at least $\sim$GHz frequencies \cite{san-paulo_detection_2007, garcia_emergence_2012}.

In summary, because the cantilever need not oscillate at $\omega_0$, it serves primarily for quasi-static positioning, while coherent actuation is provided electrically. Consequently, the AFM enters mainly through low-frequency working-distance fluctuations $\delta z(t)$ (including the contribution from AFM thermal motion) and through bias noise, both of which can be minimized in cryogenic, stiff-cantilever operation. Residual instrument vibration and drift can be treated as additional, experimentally characterizable noise sources. Quantitative estimates of the associated displacement/bias noise and their impact on coherence are provided in Sec.~S8 in the Supplemental Material \cite{SM}.

\section{Conclusions}\label{sec:conclusions}

We have presented a unified, all-mechanical scheme for coherent control and quantum-state reconstruction of the fundamental flexural mode of a suspended carbon nanotube operated in the anharmonic (Duffing/Kerr) regime. A nearby AFM tip acts as a single, localized actuator that supplies calibrated, phase-stable force waveforms for both spectroscopy and time-domain control. When the intrinsic and/or engineered anharmonicity is large compared to the relevant linewidth and control bandwidth, the lowest vibrational transition $\ket{0}\!\leftrightarrow\!\ket{1}$ can be addressed selectively, enabling effective two-level protocols such as Rabi driving and Ramsey interferometry while suppressing off-resonant excitation of higher levels. 

Within a Gorini--Kossakowski--Sudarshan--Lindblad (GKSL) framework we derived explicit expressions linking experimentally accessible signals to the key decoherence times. In particular, population dynamics and Ramsey fringes provide direct extraction of the energy-relaxation time $T_1$ and phase-coherence time $T_2$ in the mechanically defined two-level manifold, yielding a quantitative, device-level characterization of dissipation and dephasing in a mesoscopic mechanical degree of freedom. The same AFM-based calibration procedures that determine the drive strength and resonance conditions also define the operating window in which the two-level selectivity criteria are satisfied.

For full state reconstruction we developed a Wigner-tomography protocol based on displaced-parity sampling. The AFM tip implements controlled phase-space displacements $D(\alpha)$ of the CNT mode, while parity is accessed through dispersive, phonon number-dependent phase accumulation in compatible electromechanical architectures (e.g., Cooper Pair Box or cavity-based parity mapping). Scanning $\alpha$ yields $W(\alpha)$ point-by-point without global inversion, and negative regions of the Wigner function provide a direct, experimentally relevant signature of nonclassical motion whose decay enables state-resolved diagnostics of decoherence mechanisms.

We also identified the principal experimental considerations associated with AFM--CNT interactions in the intended cryogenic, noncontact operating regime. In this configuration the cantilever serves primarily for quasi-static positioning, while coherent actuation is supplied electrically via rf modulation of the tip bias (or a nearby gate), allowing the resonant \emph{AC} force component at $\omega_0$ to be set in the fN/sub-fN range required for coherent rotations even in the presence of a larger static background interaction. The dominant AFM-induced backaction channels are low frequency fluctuations of the tip-CNT working distance and bias noise, which enter as additional frequency noise (dephasing) and near-resonant force noise (residual excitation). These contributions can be quantified experimentally and kept below the intrinsic linewidth and decoherence rates under realistic operating conditions. Quantitative estimates indicate that the relevant forces and timescales lie within the reach of present cryogenic AFM and mesoscopic electromechanics. Because control and displacements are implemented via localized AFM actuation, the platform avoids optical fields and the need for dedicated on-chip microwave drive lines at the CNT, reducing added loss and heating while remaining compatible with dispersive microwave readout.

Overall, the framework provides a compact and experimentally grounded route to both spectrally selective qubit-like control of a CNT vibrational mode enabled by intrinsic/tunable anharmonicity, and phase-space tomography that accesses parity-based quantum signatures within the same device architecture. This minimal, single-actuator approach establishes a practical blueprint for quantitative studies of decoherence in mesoscopic mechanical systems and offers a pathway toward preparing and validating nonclassical motional states in CNT resonators and related nanomechanical platforms.

\section{acknowledgments}
 CS acknowledges support for this work from a Tufts Faculty Research Award (FRAC).

\section{Data Availability}
The data are not publicly available. The data are available from the authors upon reasonable request.

\bibliography{SuspendedCNPRLref_v1}

\clearpage
\onecolumngrid

\begin{center}
{\bf\large{Supplemental Material for Quantum Tomography of Suspended Carbon Nanotubes}}
\end{center}
\setcounter{section}{0}
\setcounter{subsection}{0}
\setcounter{equation}{0}
\renewcommand{\theequation}{S\arabic{equation}}
\setcounter{secnumdepth}{2} 
\renewcommand{\thesection}{S\arabic{section}}
\textbf{This PDF file includes:}

S1. System Hamiltonian for the AFM-driven CNT in the Rotating Wave Approximation

S2. GKSL Master Equation and Bloch Equations 

S3. Solutions of the Bloch Equations under Dissipation (Rabi and Ramsey)

S4. Driven response: CW carrier versus transient pulse sequences.

S5. Wigner Function for the CNT Coupled to the Environment

S6. Readout Modalities

S7. CNT--CPB Coupling Mechanism and Dispersive Readout

S8. Estimates of the AFM--CNT Interaction and Backaction Constraints

S9. Anharmonicity estimates, sideband cooling, and validity of the TLS approximation

\section{System Hamiltonian for the AFM-driven CNT in the Rotating Wave Approximation}

\textit{Physical system}. We consider a single-wall carbon nanotube (CNT) of length $L$, doubly clamped across a nanofabricated trench and positioned near an AFM tip (Fig.~1 in the main text). As discussed in the main text, the CNT fundamental flexural mode at cryogenic temperature ($\sim 10\mathrm{mK}$) can be described as an anharmonic oscillator with frequency $\omega_{0}$ and anharmonic separation $K$. When $|K|$ exceeds both the decoherence rate $\Gamma_2$ and the typical on-resonance Rabi frequency generated by AFM actuation $\Omega_R$, the driven dynamics remains confined to the lowest two states
\(\{\ket{0},\ket{1}\}\). We may therefore truncate the Hilbert space to the ground and first excited state and treat the CNT as an effective two-level system (TLS), described by the Pauli operators  \(\sigma_{z}=\ket{1}\!\bra{1}-\ket{0}\!\bra{0}\) and
\(\sigma_{x}=\ket{0}\!\bra{1}+\ket{1}\!\bra{0}\). The Hamiltonian for the undriven TLS is
\begin{equation}
H_{0}= \frac{\hbar\omega_{01}}{2}\,\sigma_{z},
\end{equation}
with $\omega_{01}$ denoting the transition frequency between the two levels. As discussed in the main text, for the Kerr model (described by Eq.~1 in the main text) the fundamental transition satisfies $\omega_{01}=\omega_0$. For notational simplicity, we therefore write $\omega_0$ for the $\ket{0}\!\leftrightarrow\!\ket{1}$ transition frequency throughout this manuscript.

\textit{Mechanical drive.}
An atomic force microscope (AFM) tip positioned near the CNT exerts a time‑dependent force
\begin{equation}
F(t)=F_{0}\cos(\omega_{d}t).
\label{eq:Fdrv}
\end{equation}
which couples to the nanotube displacement 
\(\hat x = x_{\mathrm{zpf}}(a+a^{\dagger})\).
Projecting onto the two‑level subspace gives, $\hat{x}\rightarrow x_{01}\,\sigma_x$, where $x_{01}\equiv\langle0|\hat{x}|1\rangle$ is the transition matrix element between the two states. The time–dependent drive Hamiltonian is
\begin{equation}
H_{\text{drv}}(t)=
F_{0}x_{\mathrm{01}}\cos(\omega_{d}t)\,\sigma_{x}
\equiv
\hbar g\cos(\omega_{d}t)\,\sigma_{x},
\qquad
\,g=\dfrac{F_{0}x_{\mathrm{01}}}{\hbar}.
\label{eq:gdef}
\end{equation}

\textit{Environmental bath.}
The surrounding bosonic environment is
modeled as a set of independent harmonic oscillators in thermal
equilibrium,
\(H_{B}=\sum_{k}\hbar\omega_{k}b_{k}^{\dagger}b_{k}\).
To lowest order, the CNT-environment coupling is described by the Hamiltonian \cite{breuer_theory_2009}
\begin{equation}
H_{I}= \hbar\sigma_{z}\sum_{k} g_{k}\bigl(b_{k}^{\dagger}+b_{k}\bigr),
\label{eq:HI1}
\end{equation}
which induces both energy relaxation and dephasing. The coefficients $g_{k}$ in Eq. \eqref{eq:HI1} represent coupling to bath mode $k$ with frequency $\omega_{k}$.

Collecting the above equations we obtain the \textbf{laboratory‑frame Hamiltonian}:
\begin{equation}
H_{\text{lab}}=
 \frac{\hbar\omega_{0}}{2}\,\sigma_{z}
 +\hbar g\cos(\omega_{d}t)\,\sigma_{x}
 +\sum_{k}\hbar\omega_{k}b_{k}^{\dagger}b_{k}\
 +\hbar\sigma_{z}\!\sum_{k} g_{k}(b_{k}^{\dagger}+b_{k}) .
\label{eq:Hlab}
\end{equation}

This expression represents the full Hamiltonian in the
      laboratory frame, containing a harmonically modulated
      \(\sigma_{x}\) drive and longitudinal coupling to a bosonic bath.
      
\textit{System Hamiltonian in the rotating wave approximation}. For weak driving, $g\ll \omega_0$ we transform the above Hamiltonian into a rotating frame and use the rotating wave approximation (RWA). First, introduce the unitary operator
\begin{equation}
U(t)=\exp\!\Bigl(i\frac{\omega_{d}t}{2}\sigma_{z}\Bigr),
\label{eq:Urot}
\end{equation}
which rotates the TLS about the \(z\)-axis at the drive frequency.
Operators transform according to
\(\tilde A(t)=U(t)A\,U^{\dagger}(t)\)
and by applying Eq.\,\eqref{eq:Urot} to the free and drive Hamiltonians we get:
\begin{align}
U H_{0} U^{\dagger}
      &= \frac{\hbar}{2}(\omega_{0}-\omega_{d})\,\sigma_{z}
       \equiv \frac{\hbar\Delta}{2}\,\sigma_{z},\\[6pt]
U H_{\mathrm{drv}}(t)U^{\dagger}
      &= \hbar g
         \cos(\omega_{d}t)
         \Bigl[
           \sigma_{x}\cos(\omega_{d}t)+\sigma_{y}\sin(\omega_{d}t)
         \Bigr],
\end{align}
where \(\Delta\equiv\omega_{0}-\omega_{d}\) is the detuning.
Expanding the product of cosines/sines and applying the RWA we obtain the static contribution:
\[
U H_{\mathrm{drv}}(t)U^{\dagger}\;\xrightarrow{\mathrm{RWA}}\;
 =\frac{\hbar\Omega_{R}}{2}\,
   \sigma_x,
\]
with the \emph{Rabi frequency}:
\(
\Omega_{R}=F_{0}x_{\mathrm{01}}/{\hbar}
\). Because \(U(t)\) commutes with \(H_{B}\) and \(\sigma_{z}\), the
environment Hamiltonian is \emph{unchanged}:
$\tilde H_{\mathrm{env}} = H_{B} + H_{I}$. Collecting all the terms we arrive at the \textbf{rotating‑frame
Hamiltonian}:
\begin{equation}
\tilde H =
\frac{\hbar\Delta}{2}\,\sigma_{z}
+\frac{\hbar\Omega_{R}}{2}\,\sigma_{x}
+\sum_{k}\hbar\omega_{k}b_{k}^{\dagger}b_{k}\
+\hbar\sigma_{z}\sum_{k} g_{k}\bigl(b_{k}^{\dagger}+b_{k}\bigr).
\label{eq:Hrot}
\end{equation}

\section{GKSL Master Equation and Bloch Equations}

In the previous section we have shown that after transforming to the drive rotating frame and by making the rotating–wave approximation (RWA), the system Hamiltonian in  reads (Eq.\,\eqref{eq:Hrot})
\begin{equation}
\tilde H = H_{S} + H_{B} + H_{I}
\label{eq:Htot}
\end{equation}
with
\begin{subequations}
\begin{align}
H_{S} &= \frac{\hbar\Delta}{2}\,\sigma_{z}
        +\frac{\hbar\Omega_{R}}{2}\,\sigma_{x},
        & \Delta &= \omega_{0}-\omega_{d}, \\[4pt]
H_{B} &= \sum_{k}\hbar\omega_{k}\,b_{k}^{\dagger}b_{k}, \\[4pt]
H_{I} &= \hbar\sigma_{z}\sum_{k} g_{k}\bigl(b_{k}^{\dagger}+b_{k}\bigr).
\end{align}
\end{subequations}

In the interaction picture generated by $H_{S}+H_{B}$ the density matrix
transforms as \cite{breuer_theory_2009}:
$\tilde\rho_{I}(t)=e^{i(H_{S}+H_{B})t/\hbar}\,\tilde\rho(t)\,e^{-i(H_{S}+H_{B})t/\hbar}$,
and the interaction Hamiltonian becomes:
\begin{equation}
H_{I}(t)=
\hbar\sigma_{z}(t)\,\! B(t),\quad
\begin{cases}
\sigma_{z}(t)=e^{iH_{S}t/\hbar}\sigma_{z}e^{-iH_{S}t/\hbar},\\[4pt]
B(t)=\displaystyle\sum_{k}g_{k}
     \bigl(b_{k}^{\dagger}e^{i\omega_{k}t}+b_{k}e^{-i\omega_{k}t}\bigr).
\end{cases}
\label{eq:HI}
\end{equation}

We make the following assumptions commonly employed in the theory of open quantum systems \cite{breuer_theory_2009, schlosshauer_decoherence_2007}:

\begin{enumerate}
\item \emph{Born approximation}: the composite state factorizes,
      $\tilde\rho_{I}(t)\approx\rho_{S}(t)\otimes\rho_{B}$, where
      $\rho_{B}=e^{-H_{B}/k_{B}T}/\Tr(e^{-H_{B}/k_{B}T})$ is stationary.
\item \emph{Markov approximation}: bath correlations decay much faster
      than the timescale over which $\rho_{S}(t)$ changes appreciably.
\item \emph{RWA approximation}: retain only terms
      oscillating slowly compared with the TLS transition
      frequencies.
\end{enumerate}

These assumptions lead to the following time evolution equation for the reduced density matrix \cite{breuer_theory_2009, schlosshauer_decoherence_2007}:
\begin{align}
\dot\rho_{S}(t) &=
-\frac{1}{\hbar^{2}}
\int_{0}^{\infty}\!d\tau\;
\Tr_{B}\!\Bigl[\,H_{I}(t),\,
  \bigl[H_{I}(t-\tau),\,\rho_{S}(t)\otimes\rho_{B}\bigr]\Bigr].
\label{eq:BM}
\end{align}

Next, we define the bath correlation function
\(
C(\tau)=\langle B(\tau)B(0)\rangle_{B}
\)
and write $\sigma_{z}(t)$ in the energy basis
($\hbar\Omega=\sqrt{\Delta^{2}+\Omega_{R}^{2}}$),
\(\sigma_{z}(t)=\sum_{\omega=0,\pm\Omega}\!
  e^{-i\omega t}\,A_{\omega}\),
where
\(A_{0}=(\Delta/\Omega)\sigma_{z}^{(e)}\) and
\(A_{\pm\Omega}=(\Omega_{R}/2\Omega)\sigma_{\pm}^{(e)}\).
Substituting into Eq.\,\eqref{eq:BM} and performing the $\tau$‐integral
one arrives at the standard 

\textit{Gorini–Kossakowski–Sudarshan–Lindblad (GKSL)} equation \cite{breuer_theory_2009, schlosshauer_decoherence_2007}:
\begin{equation}
\dot\rho_{S}=
-\frac{i}{\hbar}[H_{S},\rho_{S}]
+\gamma_{\downarrow}\,
  \mathcal{D}[\sigma_{-}]\rho_{S}
+\gamma_{\uparrow}\,
  \mathcal{D}[\sigma_{+}]\rho_{S}
+\gamma_{\varphi}\,
  \mathcal{D}[\sigma_{z}]\rho_{S},
\label{eq:Lindblad}
\end{equation}
with $\sigma_+=\ket{1}\!\bra{0}$, \( \sigma_- = \ket{0}\!\bra{1}\), and the \textit{dissipator}:
\begin{equation}
\mathcal{D}[\sigma]\rho_S\equiv \sigma\rho_S \sigma^{\dagger}
                      -\tfrac12\{\sigma^{\dagger}\sigma,\rho_S\}.
\label{eq:Dissipator}
\end{equation}

\textit{Physical picture --} Starting from the Hamiltonian
in Eq.\,\eqref{eq:Hrot}, the Born–Markov approximation yields the GKSL master equation
(Eq.~\eqref{eq:Lindblad}). The coefficients $\gamma_{\downarrow}$ ($\gamma_{\uparrow}$) in Eq.~\eqref{eq:Lindblad} represent the relaxation (thermal excitation) rates set by the bath spectral density at $\omega_0$, and $\gamma_{\varphi}$ accounts for pure dephasing, typically dominated by low-frequency phase noise. The system Hamiltonian $H_S$ describes the unitary evolution for the driven CNT resonator including coherent excitation by the AFM drive. The commutator
      $-i[H_{S},\rho_S]/\hbar$ in Eq.~\eqref{eq:Lindblad} is the familiar Liouville–von Neumann term
      $\partial_{t}\rho_S=-i[H_{S},\rho_S]/\hbar$ for an isolated quantum
      system. Each additional term in Eq.~\eqref{eq:Lindblad} has the structure
\(
\gamma_{j}\,\mathcal{D}[\sigma_{j}]\rho_S
\) 
 with the dissipator $\mathcal{D}[\sigma]\rho_S$ defined in Eq.\,\eqref{eq:Dissipator}.  The dissipator generates irreversible, stochastic jumps associated with processes in the environment. 
 The operators $\sigma \in \{\sigma_-,\,\sigma_+,\,\sigma_z\}$ serve as Lindblad jump operators, and capture specific physical processes (incoherent population jumps, dephasing) induced by the
      environment:

$\bullet$ $\sigma_{-}$ (\(=\ket0\!\bra1\)) :
energy relaxation ($\ket1\to\ket0$).\vspace{2pt}

$\bullet$ $\sigma_{+}$ (\(=\ket1\!\bra0\)) :
incoherent excitation ($\ket0\to\ket1$).\vspace{2pt}

$\bullet$ $\sigma_{z}$ :
pure dephasing that leaves populations unchanged.

The gain term $\sigma\rho_S \sigma^{\dagger}$ in Eq.\,\eqref{eq:Dissipator} adds population or coherence consistent with the jump. The loss term $-\tfrac12\{\sigma^{\dagger}\sigma,\rho_S\}$ subtracts
terms such that the total evolution preserves $\Tr\rho_S=1$. This competition between coherent drive and dissipative
      processes sets the visibility of Rabi oscillations, Ramsey
      fringes, and the regions where the Wigner function is negative for the CNT oscillator.

The resulting rates describe incoherent (e.g. thermal) excitation
($\gamma_\uparrow$), relaxation  ($\gamma_\downarrow$), and slow longitudinal frequency noise ($\gamma_\varphi$), thereby establishing a quantitative connection between
the microscopic properties of the environment and the observable relaxation and
dephasing times for the mechanically driven CNT resonator. These dissipative rates have dimensions of inverse time, and are set by the bath spectral density and temperature. For a transition at frequency $\omega_{0}$ \cite{breuer_theory_2009, schlosshauer_decoherence_2007}:
\begin{equation}
\gamma_{\uparrow}=2\pi J(\omega_{0})\,n(\omega_{0}),\qquad
\gamma_{\downarrow}=2\pi J(\omega_{0})\,[n(\omega_{0})+1] 
\label{eq:gamma}
\end{equation}
where $J(\omega)=2\alpha\,\omega\,e^{-\omega/\omega_c}$ is the Ohmic spectral density (with dimensionless coupling $\alpha$ and cutoff $\omega_c$), and $n(\omega)=\bigl[e^{\hbar\omega/k_BT}-1\bigr]^{-1}$ is the thermal occupation number.

\textit{Derivation of the Bloch equations from the GKSL master equation.} The Bloch vector components are defined as \cite{breuer_theory_2009}: 
\[
\,X \equiv \Tr(\sigma_{x}\rho), \qquad
        Y \equiv \Tr(\sigma_{y}\rho), \qquad
        Z \equiv \Tr(\sigma_{z}\rho)\,
\]
Using the Bloch components we can write the reduced density matrix as:
\begin{equation}
\rho_S=\tfrac12\bigl(\openone+X\sigma_{x}+Y\sigma_{y}+Z\sigma_{z}\bigr).
\label{eq:Matrix}
\end{equation}
By substituting the Bloch representation (Eq.\,\eqref{eq:Matrix}) into the GKSL equation (Eq. \eqref{eq:Lindblad}) and explicitly evaluating each term in the dissipator (Eq. \eqref{eq:Dissipator}) using the commutation relations
\([\,\sigma_{i},\sigma_{j}\,]=2i\varepsilon_{ijk}\sigma_{k}\), one obtains, after straightforward algebra, the Bloch equations for a mechanically driven carbon nanotube coupled to its environment: 
\begin{equation}
\begin{aligned}
\dot X &= -\Bigl(\tfrac12\gamma_{\downarrow}
                 +\tfrac12\gamma_{\uparrow}
                 +\gamma_{\varphi}\Bigr)X
          \;+\;\Delta\,Y, \\[6pt]
\dot Y &= -\Bigl(\tfrac12\gamma_{\downarrow}
                 +\tfrac12\gamma_{\uparrow}
                 +\gamma_{\varphi}\Bigr)Y
          \;-\;\Delta\,X
          \;+\;\Omega_{R}\,Z, \\[6pt]
\dot Z &= -(\gamma_{\downarrow}+\gamma_{\uparrow})\,Z
         \;-\;\Omega_{R}\,Y
         \;+\;(\gamma_{\uparrow}-\gamma_{\downarrow}).
\end{aligned}
\label{eq:Bloch38}
\end{equation}

All decay terms in Eq. \eqref{eq:Bloch38} are written in terms of the physical rates
\(\gamma_{\downarrow}\), \(\gamma_{\uparrow}\) (energy exchange) and
\(\gamma_{\varphi}\) (pure dephasing), which determine  the
longitudinal relaxation time \(T_{1}\) and transverse coherence time
\(T_{2}\).:

\begin{equation}
 T_{1}=\frac{1}{\gamma_{\downarrow}+\gamma_{\uparrow}},\qquad
T_{2} =\frac{1}{\dfrac{\gamma_{\downarrow}+\gamma_{\uparrow}}{2}+\gamma_{\varphi}}.   
\end{equation}

The longitudinal relaxation time $T_{1}$ characterizes population relaxation toward thermal equilibrium, encompassing both energy decay from $\ket{1}$ to $\ket{0}$ and thermal repopulation of the excited state. Physically, $T_{1}$ quantifies how rapidly the excited nanotube state $\ket{1}$ relaxes to $\ket{0}$ at rate $\gamma_{\downarrow}$, while the thermal bath repopulates $\ket{1}$ at rate $\gamma_{\uparrow}$. The phase-coherence time $T_{2}$ governs the exponential decay of the off-diagonal density-matrix elements, i.e., the decay of coherence between $\ket{0}$ and $\ket{1}$. In the relaxation-limited regime, where $\gamma_{\varphi}\approx 0$, one has $T_{2}\simeq 2T_{1}$. By  defining the thermal inversion 
\(Z_{\!\mathrm{eq}}=-\tanh\!\bigl(\hbar\omega_{0}/2k_{B}T\bigr)
  =(\gamma_{\uparrow}-\gamma_{\downarrow})/
    (\gamma_{\uparrow}+\gamma_{\downarrow})\), the Bloch equations 
Eqs.~\,\eqref{eq:Bloch38} take the form given in the main text:

\begin{subequations}\label{eq:Bloch}
\begin{align}
\dot X &= -\Gamma_{2}\,X + \Delta Y, \tag{\ref{eq:Bloch}a}\\
\dot Y &= -\Gamma_{2}\,Y -\Delta X + \Omega_{R}\,Z, \tag{\ref{eq:Bloch}b}\\
\dot Z &= -\Gamma_{1}\,(Z-Z_{eq}) - \Omega_{R}\,Y, \tag{\ref{eq:Bloch}c}
\end{align}
\end{subequations}
with rates $
\Gamma_{1}=\gamma_{\downarrow}+\gamma_{\uparrow}, \
\Gamma_{2}=\frac{\Gamma_{1}}{2}+\gamma_{\varphi}$. The Rabi frequency for the mechanically driven CNT is  
\(\Omega_{R}=F_{0}x_{\mathrm{zpf}}/\hbar\).

\textit{In summary}, starting from the Hamiltonian
in Eqs.\,\eqref{eq:Htot}-\eqref{eq:HI}, the Born–Markov approximation yields the GKSL master equation
Eq.~\eqref{eq:Lindblad}.
The resulting rates describe mechanical damping (\(\gamma_{\downarrow}\)),
thermal excitation (\(\gamma_{\uparrow}\)) and low‑frequency noise
(\(\gamma_{\varphi}\)), thereby establishing a quantitative connection between
the microscopic properties of the environment and the observable relaxation and
dephasing times for the mechanically driven CNT resonator.

\section{Solutions of the Bloch Equations under Dissipation (Rabi and Ramsey)}

In this section we show that on resonance ($\Delta=0$), at dilution temperatures ($\sim10$ mK), and with the system initialized in the ground state $\ket{0}$, the mechanically driven CNT  undergoes damped Rabi oscillations. By choosing zero detuning \(\Delta=\omega_{0}-\omega_{d}=0\) and shift \(z(t)=Z(t)-Z_{\!\mathrm{eq}}\),  
the Bloch equations (\ref{eq:Bloch}) become:
\begin{subequations}\label{eq:hom}
\begin{align}
\dot X &= -\Gamma_{2}\,X, \\[2pt]
\dot Y &= -\Gamma_{2}\,Y + \Omega_{R}(z+Z_{\!\mathrm{eq}}), \\[2pt]
\dot z &= -\Gamma_{1}\,z - \Omega_{R}\,Y,
\end{align}
\end{subequations}
which are homogeneous in \((Y,z)\) apart from the constant term
\(\Omega_{R}Z_{\!\mathrm{eq}}\).

After eliminating $X$ and $Y$, the shifted $Z$–component satisfies the following equation:
\begin{equation}
  \ddot z + 2\lambda\dot z + (\lambda^{2}+\widetilde \Omega^{2})z = -\,\Omega_{R}^{2}Z_{\text{eq}},
  \label{eq:forced}
\end{equation}

with
\[\lambda\equiv\tfrac{\Gamma_{1}+\Gamma_{2}}{2}, \quad
  \widetilde\Omega\equiv\sqrt{\Omega_{R}^{2}-\bigl(\tfrac{\Gamma_{1}-\Gamma_{2}}{2}\bigr)^{2}},\]

\noindent
The general solution of Eq. \eqref{eq:forced} is:
\begin{equation}
  z(t)=- \frac{\Omega_{R}^{2}}{\lambda^{2}+\widetilde\Omega^{2}}Z_{\text{eq}}+e^{-\lambda t}\bigl[C_{1}\cos(\widetilde\Omega t)+C_{2}\sin(\widetilde \Omega t)\bigr].
  \label{eq:zgen}
\end{equation}
The physical $Z$–Bloch component is:
\begin{equation}
  Z(t)=z(t)+Z_{\text{eq}}.
  \label{eq:Zdef}
\end{equation}
Once $Z(t)$ is known, the probability of occupying the excited vibrational state $\ket{1}$ is given by \cite{breuer_theory_2009}:
\begin{equation}
  \;P_{1}(t)=\frac{1+Z(t)}{2}\;, \quad \bigl(P_{0}(t)=1-P_{1}(t)=\tfrac{1-Z(t)}{2}\bigr),
  \label{eq:P1general}
\end{equation}

\textit{Solutions for initial ground state $\ket{0}$.} We assume the CNT is initially in the ground state  
\(\ket{\psi(0)}=\ket0\), so the Bloch vector starts at the south pole:
\begin{equation}
  Z(0)=-1 \;\Longrightarrow\; z(0)=-1-Z_{\text{eq}}, \qquad \dot z(0)=\Gamma_{1}(1+Z_{\text{eq}}).
  \label{eq:initB}
\end{equation}

\noindent
Moreover, we assume that the system is at cryogenic temperatures ($\sim10$mK) such that the thermal inversion $Z_{\!\mathrm{eq}}=\tanh\!\bigl(\hbar\omega/2k_{B}T\bigr)\approx-1$, as discussed in the main text. By substituting these conditions into the general solution in Eq. \eqref{eq:zgen}, we obtain the expression for the excited-state occupation probability given in the main text:
\begin{equation}
P_1(t)=\frac{\Omega_R^2}{2\big(\Omega_R^2+\Gamma_1\Gamma_2\big)}\Big[1-e^{-(\Gamma_1+\Gamma_2)t/2}\Big(\cos(\widetilde{\Omega} t)+\frac{\Gamma_1-\Gamma_2}{2\,\widetilde{\Omega}}\sin(\widetilde{\Omega} t)\Big)\Big],
\label{eq:P1}
\end{equation}

\textit{Ramsey oscillations.} Assume the CNT starts in the ground state \(\ket{\psi(0)}=\ket0\), with Bloch vector
$(0,0,-1)$. Consider two short AFM $\pi/2$ pulses separated by a free-evolution interval $\tau$, and work in the short-pulse regime $t_{\pi/2}=\frac{\pi}{2\Omega_R}\ll \tau,\ T_1,\ T_2$,
so dissipation during the pulses is negligible.
Immediately after the first pulse the CNT is in the state $\ket{\psi(0)}=\tfrac{1}{\sqrt{2}}\big(\ket{0}+\ket{1}\big)$. We choose the phase of the first resonant $\pi/2$ pulse such that the Bloch vector becomes: $(1,0,0)$. During the free interval $\tau$ the drive is off, $\Omega_R(t)=0$, and the Bloch equations (Eqs.~\ref{eq:Bloch}) decouple. With initial condition $(X,Y)=(1,0)$ at $t=0$ one obtains
\begin{align}
X(\tau) &= e^{-\tau/T_2}\cos(\Delta\tau),\label{eq:Xfree}\\
Y(\tau) &= e^{-\tau/T_2}\sin(\Delta\tau),\label{eq:Yfree}
\end{align}
independent of $T_1$. The longitudinal component $Z(\tau)$ does not enter the ideal Ramsey signal below.

Let the second resonant $\pi/2$ pulse have phase $\phi_0$ (set by the programmed phase between the two AFM pulses). A $\pi/2$ rotation about an equatorial axis with azimuth $\phi_0$ maps the equatorial projection $(X,Y)$ onto the $Z$ axis according to
\begin{equation}
Z_{\mathrm{f}}=X(\tau)\cos\phi_0+Y(\tau)\sin\phi_0,
\label{eq:Zmap}
\end{equation}
and the excited-state probability after the second pulse is
\begin{equation}
P_1^{\mathrm{R}}(\tau)=\frac{1+Z_{\mathrm{f}}}{2}.
\label{eq:PeDef}
\end{equation}
Substituting Eqs.~\eqref{eq:Xfree}--\eqref{eq:Yfree} into Eq.~\eqref{eq:Zmap} and then into Eq.~\eqref{eq:PeDef} yields the expression for the Ramsey excited-state probability given in the main text:
\begin{equation}
P_1^{\mathrm{R}}(\tau)=\tfrac{1}{2}\Big[1+e^{-\tau/T_2}\cos\!\big(\Delta\tau+\phi_0\big)\Big].
\label{eq:PR}
\end{equation}

Eq.~\eqref{eq:PR} shows that the fringe frequency is set by the detuning $\Delta$, and the contrast by $T_2$. The energy-relaxation time $T_1$ does not enter explicitly because the $\pi/2$ mapping eliminates $Z(\tau)$.
In summary, Eqs.(~\ref{eq:Bloch}), (\ref{eq:P1}), \eqref{eq:PR}  together characterize energy relaxation and dephasing of the mechanically driven CNT, linking bath
parameters to experimentally measurable Rabi‑ and Ramsey‑decay envelopes.

\section{Driven response: CW carrier versus transient pulse sequences.}

As discussed in the main text, we distinguish between the \emph{CW carrier} used to define the near-resonant drive and the \emph{time-domain} operation of the control and measurement protocols. For spectroscopy and for defining the rotating-frame Hamiltonian, we model the applied force as a continuous-wave (CW) drive,
$F(t)=F_{0}\cos(\omega_{d}t)$, which yields the standard RWA form and identifies the Rabi scale $\Omega_{R}\sim F_{0}x_{01}/\hbar$. Operationally, however, the Ramsey and tomography protocols are intrinsically time-domain: the CNT is driven in short, phase-coherent segments (``pulses'') separated by free evolution and readout, rather than being interrogated via the steady-state response of a continuously driven resonator. These protocols could be implemented by \emph{gating} the CW carrier with a finite envelope,
\begin{equation}
F(t)=F_{0}\,s(t)\cos(\omega_{d}t+\phi),
\end{equation}
where $s(t)$ is a square or smoothly shaped window. Therefore the relevant dynamics is the \emph{transient} response during drive segments of duration $t_{p}$ interleaved with free-evolution intervals, rather than the long-time steady-state amplitude of an indefinitely driven oscillator. 

In the parameter ranges of Table I in the main text, with $Q=10^{4}$), the energy-decay time is $T_{1}=Q/\omega_{0}$, giving $T_1\simeq 0.29~\mu\mathrm{s}$, $7.4~\mu\mathrm{s}$, and $29.6~\mu\mathrm{s}$ for $L=100$, $500$, and $1000$~nm, respectively. Accordingly, the pulse sequences can be chosen so that each gated drive segment remains in the transient regime: for the GHz device we use $t_{p}\sim 0.1~\mathrm{\mu s}$ (so $t_{p}/T_{1}\simeq 0.34$), while for the lower-frequency devices we use $t_{p}\sim 1~\mu\mathrm{s}$ (so $t_{p}/T_{1}\simeq 0.14$ at $221$~MHz and $t_{p}/T_{1}\simeq 0.034$ at $54$~MHz). These gated CW bursts also contain many carrier cycles at $\omega_d\simeq\omega_{0}$, namely $N_{\mathrm{cyc}}=t_{p}f_{d}\approx 5.4\times 10^{2}$ (GHz device, $t_p=0.1~\mathrm{\mu s}$), $\approx 2.2\times 10^{2}$ (221~MHz device, $t_p=1~\mu$s), and $\approx 54$ (54~MHz device, $t_p=1~\mu$s), so they are spectrally narrow around $\omega_{d}$ and are appropriately treated as near-resonant bursts rather than impulsive kicks.

For completeness, we note that the steady-state classical forced-response description under \emph{indefinite} CW driving is characterized by the frequency ratio $r=\omega_{d}/\omega_{0}$, damping ratio $\zeta \approx 1/(2Q)$, and magnification factor
\begin{equation}
M(r,\zeta)=\frac{1}{\sqrt{(1-r^{2})^{2}+(2\zeta r)^{2}}},
\end{equation}
which sets the ring-up amplitude and hence the importance of drive-induced geometric nonlinearity (tension) in the classical steady state. In the present work, however, the experimentally relevant control parameter is the pulse area (equivalently the rotation angle set by $\Omega_{R}$ and $t_{p}$ in the effective two-level description), while any steady-state magnification is not the operative quantity for the Ramsey and tomography sequences.

\section{Wigner Function for the CNT Coupled to the Environment}
In this subsection we calculate the Wigner function explicitly for the CNT flexural mode. As established in the previous sections, immediately after preparation the CNT is in the coherent superposition
\(\ket{\psi(0)}=(\ket0+\ket1)/\sqrt2\) with density matrix
\[
\rho(0)=\frac12
\begin{pmatrix}
1 & 1\\[2pt]
1 & 1
\end{pmatrix}
\]

Coupling to a bath produces longitudinal relaxation at rate
\(\Gamma_1=1/T_1\) and transverse decay at rate \(\Gamma_2=1/T_2\).
Within the Born–Markov approximation, and in the relaxation-limited regime
(negligible pure dephasing, such that \(T_2\simeq2T_1\)) and cold-bath limit
(\(n_{\mathrm{th}}\!\ll\!1\)), the density-matrix elements evolve as \cite{breuer_theory_2009}:
\begin{align}
\rho_{11}(t) &= \rho_{11}(0)\,e^{-\Gamma_1 t}, &
\rho_{00}(t) &= 1-\rho_{11}(t), \\[4pt]
\rho_{01}(t) &= \rho_{01}(0)\,e^{-t/T_2}\,e^{-i\omega_{0}t}
              = \rho_{01}(0)\,e^{-(\Gamma_1/2)t}\,e^{-i\omega_{0}t}, &
\rho_{10}(t) &= \rho_{01}^{\ast}(t).
\label{eq:rho}
\end{align}

The Wigner function for a state
\(\rho=\sum_{m,n=0}^{1}\rho_{mn}\ket m\!\bra n\) is \cite{breuer_theory_2009}:
\begin{equation}
W(\alpha)=\frac{2}{\pi}\,e^{-2|\alpha|^{2}}
\Bigl[
\rho_{00}
+\rho_{11}\bigl(4|\alpha|^{2}-1\bigr)
+2\,\Re\!\bigl(2\alpha\,\rho_{10}\bigr)
\Bigr],
\label{eq:Wskeleton}
\end{equation}
where \(\alpha=(X+iP)/\sqrt2\) is the phase-space coordinate. Writing \(\alpha=r\,e^{i\theta}\) with \(r\ge0\), and
inserting ~\eqref{eq:rho} into Eq.~\eqref{eq:Wskeleton} yields the time-dependent Wigner function:
\begin{equation}
W(r,\theta;t)=\frac{2}{\pi}\,e^{-2r^{2}}
\Bigl[
\bigl(1-P(t)\bigr)
+P(t)\,\bigl(4r^{2}-1\bigr)
+2\,r\,e^{-t/T_{2}}\cos\!\bigl(\theta-\omega_{0}t\bigr)
\Bigr],
\end{equation}
with \(P(t)=\rho_{11}(t)=\tfrac12\,e^{-t/T_{1}}\).
\\

\textbf{Limiting cases:}
\begin{itemize}
\item \emph{Initial state \((t=0)\):}\quad
\(P(0)=\tfrac12\)\(e^{-t/T_1}=1/2\),
\[
W(r,\theta;0)=\frac{2}{\pi}e^{-2r^{2}}
\Bigl[\,2r^{2}+2r\cos\theta\Bigr],
\]
showing a negative lobe near the origin (quantum-interference signature shown in panels (a), (d) in Fig. 3 in the main text).
\item \emph{Intermediate times ($t\sim T_1$):}\quad
the coherence term decays as \(e^{-t/T_2}\), and the system is approaching a positive mixture of \(\ket0\) and \(\ket1\) (panels (b), (e) in Fig. 3 in the main text)).
\item \emph{Long times \((t\gg T_{1})\):}\quad
\(P(t)\to0\) so \(e^{-t/T_2}\to0\),
\(
W\to \dfrac{2}{\pi}e^{-2r^{2}},
\)
the Gaussian Wigner function of the ground state \(\ket 0\) (panels (c), (f) in Fig. 3 in the main text)).
\end{itemize}

\section{Readout modalities}
We propose four different readout schemes: AFM deflection, Dispersive CPB, Dispersive cavity, and Direct mechanics cavity. All four methods share the same all-mechanical basic controls (AFM $\pi/2$ rotations and displacements), extending the Ramsey interferometer into a full, state-resolved characterization of the phase space.
\subsection{AFM deflection and direct population readout} 

Immediately after a control sequence (Rabi/Ramsey or displacement), the CNT displacement is transduced by the AFM probe (optical/electrical deflection or weak probe force) described by
\begin{equation}
H_{\rm meas}(t) = -F_{\rm probe}(t)\,\hat{x} = -F_{\rm probe}(t)\,x_{\rm zpf}(a+a^\dagger).
\label{A1}
\end{equation}
The resulting demodulated AFM signal provides access to a mechanical quadrature (and hence to the coherent response) at the drive frequency, and the linear demodulation of the quadrature distinguishes $\ket{1}$ from $\ket{0}$, returning $P_1$. In practice, the mechanical ringdown of the demodulated signal yields the longitudinal relaxation time $T_1$, while sweeping the Ramsey delay $\tau$ gives the detuning frequency $\Delta$ and phase coherence time $T_2$. This channel is thus used to measure population and characterize decoherence. Practical implementation follows the Ramsey calibration: (i) set $\omega_d\approx\omega_{0}$ and determine $\Omega_R$ via a Rabi scan; (ii) choose $t_{\pi/2}=\pi/(2\Omega_R)$, (iii) synchronize two identical $\pi/2$ pulses with programmable delay $\tau$, and (iv) demodulate the AFM signal during a short integration window after the second pulse. Typical operating times are $t_{\pi/2}\sim 0.1$–$0.3\,\mu$s, $\tau=0.1$–$50\,\mu$s. Readout is performed via lock-in or IQ demodulation referenced to the drive frequency $\omega_d$ \cite{lee_strong_2023, wollack_quantum_2022}. 
\subsection{Dispersive CPB readout and parity mapping} A Cooper–pair box dispersively coupled to the CNT implements a phase proportional to the phonon-number operator $\hat n=a^\dagger a$ during the Ramsey free-evolution window \cite{armour_entanglement_2002, blencowe_quantum_2004}. The Jaynes–Cummings interaction is
\begin{equation}
\frac{H}{\hbar}=\omega_{0}\,a^\dagger a+\frac{\omega_{q}}{2}\sigma_{z}
+g_{m}\big(a\sigma_{+}+a^\dagger\sigma_{-}\big),
\label{B1}
\end{equation}
where $\omega_{0}$ is the CNT flexural-mode frequency, $\omega_q$ is the CPB qubit transition frequency, and $g_m$ is the CPB–CNT coupling. The CPB–CNT mechanical-mode detuning is $\Delta_q\equiv\omega_q-\omega_{0}$. For $|\Delta_q|\gg g_m$, the interaction reduces to the dispersive Hamiltonian \cite{blais_cavity_2004,blais_circuit_2021,boissonneault_dispersive_2009,zueco_qubit_2009}:
\begin{equation}
\frac{H_{\rm int}^{(\rm CPB)}}{\hbar}\simeq 
\chi\,a^\dagger a\,\frac{\sigma_z}{2},\qquad 
\chi\simeq \frac{2g_m^{2}}{\Delta_q}\ .
\label{B2}
\end{equation}

Mechanical implementations of such dispersive couplings have been demonstrated with CPB/transmon devices \cite{lahaye_nanomechanical_2009, pirkkalainen_cavity_2015, lee_strong_2023}.
During a calibrated window $t_\pi=\pi/\chi$, the CPB acquires the number-dependent phase $\chi t_{\pi}\,\hat n$, implementing $e^{i\pi\hat n}$. When the Ramsey sequence is preceded by a mechanical displacement $D(\alpha)$ and performed with the pulse sequence $\{\pi/2,\,t_\pi,\,-\pi/2\}$ (where $-\pi/2$ denotes a $\pi/2$ rotation with a $\pi$-phase flip of the drive), the excited-state probability encodes the displaced parity signal as given in Eq.(22) in the main text. High-contrast mapping requires \(2\pi/\omega_{0}\ll t_{\pi}\ll T_{2}\), and dephasing during \(t_{\pi}\) limits the contrast. We note that a weak Duffing (self-Kerr) term arising from tension-induced geometric effects and/or electrostatic forces \cite{sazonova_tunable_2004, garcia-sanchez_mechanical_2007, huttel_nanoelectromechanics_2008, zueco_qubit_2009, eichler_nonlinear_2011, wang_creating_2016, semiao_kerr_2009, lifshitz_cross_nonlinear_2009, rhoads_nonlinear_2010}
produces nonlinear phase evolution $U_K(t)=\exp[-i(K/2)\hat n(\hat n-1)t]$, which can modify the Wigner distribution during the displacement and mapping windows. Implementing the parity gate $\exp(i\pi\hat n)$ requires a dispersive number-dependent phase shift linear in $\hat n$ (e.g., via a CPB or cavity \cite{lahaye_nanomechanical_2009, pirkkalainen_cavity_2015, lee_strong_2023, armour_entanglement_2002, aspelmeyer_cavity_2014}). Detailed calculations of the CNT-CPB coupling are provided in Section S6 below.

\subsection{Dispersive cavity readout} A microwave cavity operated in the dispersive regime provides phase-sensitive readout of the CPB qubit state during the Ramsey sequence. The effective readout Hamiltonian is 
\begin{equation}
\frac{H_{\rm read}}{\hbar}
= \omega_c\, a_c^\dagger a_c
+ \frac{\tilde\omega_q}{2}\,\sigma_z
+ \chi_c\, a_c^\dagger a_c\,\frac{\sigma_z}{2}.
\label{C1}
\end{equation}
Here, $\omega_c$ is the bare cavity angular frequency; $a_c$ ($a_c^\dagger$) annihilates (creates) a cavity photon; $\tilde\omega_q$ is the dressed CPB qubit transition frequency (including static Lamb/AC-Stark shifts); and $\chi_c$ is the dispersive coupling (``cavity pull'') that produces a CPB state–dependent cavity shift $\omega_c \!\to\! \omega_c \pm \chi_c/2$ for $\sigma_z=\pm1$ \cite{blais_cavity_2004,blais_circuit_2021,boissonneault_dispersive_2009,zueco_qubit_2009}. Further details of the dispersive coupling are given in Section S6 below. 

\emph{Homodyne detection --} The cavity is driven near $\omega_c$ and the transmitted (or reflected) field is routed through cryogenic circulators/isolators and a high electron mobility transistor (HEMT) low-noise amplifier \cite{cui_feedback_2013, riste_initialization_2012}. The signal is mixed with a local oscillator (LO) phase-locked to the cavity probe in an IQ mixer to yield baseband voltages $I(t)$ and $Q(t)$. The instantaneous phase $\phi(t)=\arctan[Q(t)/I(t)]$ (or an appropriate quadrature) tracks the CPB qubit observable $\sigma_z$, since the CPB state–dependent pull $\pm\chi_c/2$ rotates the cavity response in the complex plane \cite{cui_feedback_2013, riste_initialization_2012}. Integration over a window set by the cavity linewidth provides the Ramsey or parity-mapped population used in the analysis.

\subsection{Direct mechanics cavity readout} As an alternative to CPB-based readout, one may couple the cavity directly to the CNT mode to produce either a cross-Kerr interaction \cite{aspelmeyer_cavity_2014}:
\begin{equation}
\frac{H_{\rm int}^{(\rm mc)}}{\hbar}\simeq \chi_{mc}\, a^\dagger a\, a_c^\dagger a_c,
\label{C2}
\end{equation}
or a radiation-pressure interaction \cite{blais_circuit_2021, aspelmeyer_cavity_2014}:
\begin{equation}
\frac{H_{\rm rp}}{\hbar}= -g_0\, a_c^\dagger a_c\,(a+a^\dagger),
\label{C3}
\end{equation}
where $a, a^\dagger$ denote the CNT mechanical modes and $a_c$, $a_c^\dagger$ the cavity modes.
During the Ramsey free-evolution window, these couplings generate an effective phase rate $\chi_{\rm eff}$ proportional to the CNT phonon number $\hat n$, so the same parity gate $e^{i\pi\hat n}$ is realized by choosing $t_\pi=\pi/\chi_{\rm eff}$. The tomography framework and normalization are the same as described above.

\subsection{Practical timing, calibration, and acquisition}
Mechanical rotations are kept short \((t_{\pi/2}\!\ll\!\tau,\,T_2)\) to minimize pulse-induced dephasing. The relative phase of the second \(\pi/2\) rotation is set either by a \(\pi\) phase flip of the drive or by a half-period timing slip \(t=\pi/\omega_d\). The parity-mapping window \(t_\pi\) is chosen using an estimate of \(\chi\) and then fine-tuned by maximizing contrast, subject to the condition \(2\pi/\omega_0 \ll t_\pi \ll T_2\) (see part (b) above). The Ramsey population \(P_1^R(\tau)\) is fitted to the expressions given in Eq.~\eqref{eq:PR}. 

Calibration follows the sequence summarized in (A): Rabi scans set \(\Omega_R\) and \(t_{\pi/2}\), Ramsey scans determine \(\Delta\) and \(T_2\), and mechanical ringdown yields \(T_1\). The displacement for \(D(\alpha)\) is obtained from the small-signal transfer function.
The waveform generator, digitizer, and LO share a common frequency reference. Hardware gating minimizes relative jitter and slow LO-phase drift is tracked with interleaved references \cite{cui_feedback_2013, riste_initialization_2012}.

 For cavity readout, the I/Q record (subsection (C) above) is integrated with a matched filter over a window on the order of the cavity photon lifetime, chosen to remain well within the CPB \(T_1\). For AFM deflection readout, the demodulated quadrature at \(\omega_d\) is integrated over a short \(\mu\)s-scale window placed after the second pulse to avoid drive transients, while remaining much shorter than $T_1$.

\section{CNT--CPB Coupling Mechanism and Dispersive Readout}

This section provides (i) an explicit circuit-level origin of the nanotube--Cooper-pair-box (CPB) coupling, (ii) the connection between that coupling and the Jaynes--Cummings/dispersive forms used in the preceding sections, and (iii) a concrete measurement chain and pulse sequence showing how the CPB readout encodes phonon-number parity (and hence supports Wigner tomography). Our discussion follows standard CPB--mechanical coupling treatments in which a mechanical electrode modulates the CPB gate charge and thereby produces a longitudinal (charge-basis) coupling \cite{armour_entanglement_2002,blencowe_quantum_2004}. In particular, Ref.~\cite{blencowe_quantum_2004} gives an explicit circuit, Hamiltonian, control sequence, and rf-SET charge readout for a mechanically coupled CPB, including the canonical interaction term linear in displacement and proportional to a CPB Pauli operator \cite{blencowe_quantum_2004}. Ref.~ \cite{blais_circuit_2021} gives a modern circuit-QED perspective on dispersive readout and gate-based measurements.

\subsection{Physical origin of the CNT--CPB coupling (capacitive gate-charge modulation)}
\label{SM:origin}

A CPB consists of a small superconducting island connected to a reservoir through a Josephson junction (capacitance $C_J$, Josephson energy $E_J$) and biased by one or more gate capacitors. In the charge regime, the electrostatic energy is set by the charging energy $E_C=e^2/(2C_\Sigma)$, where $C_\Sigma$ is the total island capacitance. If one of the gate capacitances is formed (or significantly affected) by the nearby CNT, then the CNT displacement $x$ modulates that capacitance and hence the dimensionless gate charge $n_g(x)$, producing a coupling between the CPB and the CNT mode.

Concretely, let the CNT act as (or effectively modulate) a gate electrode with capacitance $C_m(x)$ biased at voltage $V_m$. Then
\begin{equation}
n_g(x)=\frac{1}{2e}\Bigl[C_g V_g + C_m(x)\,V_m + \cdots\Bigr],
\qquad
C_m(x)\simeq C_m^{(0)} + \left.\frac{dC_m}{dx}\right|_{0} x ,
\end{equation}
so that to leading order
\begin{equation}
\delta n_g(x)\equiv n_g(x)-n_g(0)=\left(\frac{V_m}{2e}\frac{dC_m}{dx}\right)x \equiv \eta\,x .
\end{equation}
In the standard two-charge-state approximation near the charge degeneracy point, the CPB Hamiltonian takes the qubit form \cite{blencowe_quantum_2004}:
\begin{equation}
H_{\rm CPB}\;=\;4E_C\,\delta n\,\sigma_z \;-\;\frac{E_J}{2}\sigma_x,
\qquad
\delta n \equiv n_g-\Bigl(n+\frac{1}{2}\Bigr),
\end{equation}
and the mechanical mode is $H_m=\hbar\omega_0 a^\dagger a$ with displacement operator $x=x_{\rm zpf}(a+a^\dagger)$.
Expanding $\delta n=\delta n_0+\eta x$ gives the interaction
\begin{equation}
H_{\rm int}=\hbar\lambda\,(a+a^\dagger)\sigma_z,
\qquad
\lambda=\frac{4E_C}{\hbar}\,\eta\,x_{\rm zpf}
=\frac{4E_C}{\hbar}\left(\frac{V_m}{2e}\frac{dC_m}{dx}\right)x_{\rm zpf}.
\label{SM:longitudinal}
\end{equation}
This is the standard electromechanical coupling mechanism for charge qubits: mechanical motion modulates the CPB gate charge and enters as a longitudinal coupling in the charge basis. An explicit version of this Hamiltonian, including the $\propto (a+a^\dagger)\sigma_z$ term and the relevant circuit elements, is written down in Ref.~\cite{blencowe_quantum_2004} (their ``Cooper box--cantilever'' model) \cite{blencowe_quantum_2004}. The same reference also provides an explicit expression for the coupling constant in terms of circuit parameters, the zero-point motion, and the electrode gap \cite{blencowe_quantum_2004}.

\subsection{From longitudinal coupling to Jaynes--Cummings form and the dispersive interaction}
\label{SM:JCdispersive}

To connect Eq.~\eqref{SM:longitudinal} to the Jaynes--Cummings form used in the main text, one moves from the CPB charge basis to the qubit energy eigenbasis. Writing the qubit Hamiltonian as
\begin{equation}
H_q=\frac{\hbar\omega_q}{2}\,\tilde\sigma_z,
\qquad
\hbar\omega_q=\sqrt{(8E_C\delta n_0)^2+E_J^2},
\end{equation}
the operator $\sigma_z$ (charge basis) becomes a linear combination of $\tilde\sigma_z$ and $\tilde\sigma_x$ (energy basis),
\begin{equation}
\sigma_z=\cos\theta\,\tilde\sigma_z+\sin\theta\,\tilde\sigma_x,
\qquad
\tan\theta=\frac{E_J}{8E_C\delta n_0}.
\end{equation}
Thus the interaction contains both longitudinal and transverse components in the energy basis:
\begin{equation}
H_{\rm int}=\hbar\lambda(a+a^\dagger)\bigl(\cos\theta\,\tilde\sigma_z+\sin\theta\,\tilde\sigma_x\bigr).
\end{equation}
Near the charge degeneracy point ($\delta n_0\to 0$), one has $\theta\to \pi/2$ and the coupling is predominantly transverse, enabling exchange interactions. In the rotating-wave approximation (RWA), the transverse term yields the Jaynes--Cummings coupling
\begin{equation}
H_{\rm JC}=\hbar\omega_0 a^\dagger a+\frac{\hbar\omega_q}{2}\tilde\sigma_z
+\hbar g_m\,(a\,\tilde\sigma_+ + a^\dagger \tilde\sigma_-),
\qquad
g_m \equiv \lambda\sin\theta.
\label{SM:JC}
\end{equation}
This is the interaction quoted in the section S.6 (Eq.~\eqref{B1}). In the dispersive regime $|\Delta_q|\equiv|\omega_q-\omega_0|\gg g_m$, second-order perturbation theory (or a Schrieffer--Wolff transformation) gives the number-dependent phase shift of the qubit (and equivalently a qubit-state-dependent shift of the mechanical frequency),
\begin{equation}
H_{\rm disp}\simeq \hbar\omega_0 a^\dagger a + \frac{\hbar}{2}\bigl(\omega_q+\chi\bigr)\tilde\sigma_z
+\hbar\chi\,a^\dagger a\,\frac{\tilde\sigma_z}{2},
\qquad
\chi \simeq \frac{2g_m^2}{\Delta_q},
\label{SM:dispersive}
\end{equation}
consistent with Eq.~(B2) in the main text (up to conventional factors depending on the precise definition of $\chi$ and the retained Lamb shifts) \cite{blais_cavity_2004,blais_circuit_2021,boissonneault_dispersive_2009,zueco_qubit_2009}.

\subsection{How the measurement is performed: CPB Ramsey readout and parity mapping}
\label{SM:parity}

\paragraph{(a) Physical readout of the CPB.}
The CPB charge state can be measured with high sensitivity using an rf single-electron transistor (rf-SET) capacitively coupled to the CPB island \cite{armour_entanglement_2002,blencowe_quantum_2004}. In Ref.~\cite{blencowe_quantum_2004}, the rf-SET readout stage is described explicitly: after the control/interaction window, the rf-SET is biased to produce a tunneling current whose value depends on the CPB island charge \cite{blencowe_quantum_2004}. That reference also discusses the requirement that the measurement time be short compared to the CPB state lifetime set by backaction and voltage-noise fluctuations, and provides representative numbers illustrating that this condition can be satisfied in practice \cite{blencowe_quantum_2004}.

\paragraph{(b) Number-dependent phase accumulation and the parity gate.}
In the dispersive regime, Eq.~\eqref{SM:dispersive} implies that during a free-evolution interval of duration $t$ the joint evolution includes the unitary
\begin{equation}
U_{\rm int}(t)=\exp\!\left[-i\frac{\chi t}{2}\,a^\dagger a\,\tilde\sigma_z\right].
\end{equation}
If the CPB is prepared in a superposition of $\tilde\sigma_z$ eigenstates, the mechanical Fock components $|n\rangle$ imprint phases $\propto n$ on the CPB, i.e., the CPB phase becomes conditioned on the phonon number. Choosing the calibrated interaction time
\begin{equation}
t_\pi=\frac{\pi}{\chi},
\label{SM:tpi}
\end{equation}
yields a controlled phase of $\pi n$ between the CPB components, which is precisely the parity phase factor $(-1)^n=e^{i\pi n}$.

\paragraph{(c) Ramsey-to-parity mapping (and displaced parity for Wigner tomography).}
A standard way to convert this number-dependent phase into an experimentally accessible population is a Ramsey sequence on the CPB. The essential pulse pattern is
\begin{equation}
\left\{R_{\pi/2}\right\}\;\; \xrightarrow{\;\;t_\pi\;\;}\;\; \left\{R_{-\pi/2}\right\}\;\;\xrightarrow{\;\;\text{CPB readout}\;\;}\;\; P_e,
\end{equation}
where $R_{\pm\pi/2}$ are $\pm\pi/2$ rotations of the CPB about an equatorial axis of its Bloch sphere. Because $U_{\rm int}(t_\pi)$ converts the mechanical number $n$ into a CPB phase $\pi n$, the final excited-state probability $P_e$ becomes proportional to the mechanical parity expectation value $\langle (-1)^{\hat n}\rangle$ (up to an overall contrast factor set by CPB coherence during $t_\pi$). This logic is closely related to the CPB pulse/interference protocol explicitly analyzed in Ref.~\cite{blencowe_quantum_2004}, where a controlled evolution period imprints mechanical information onto CPB charge-state probabilities that are then read out \cite{blencowe_quantum_2004}. (In our application, the dispersive regime is used so that the imprinted phase is \emph{number dependent} rather than producing large conditional displacements.)

To obtain the full Wigner function $W(\alpha)$ one measures \emph{displaced parity}. As discussed in the main text, this is achieved by applying a mechanical displacement $D(\alpha)$ prior to the parity-mapping Ramsey sequence:
\begin{equation}
W(\alpha)=\frac{2}{\pi}\,\Tr\!\left[D(-\alpha)\rho_s\,D(\alpha)\,(-1)^{\hat n}\right].
\end{equation}
Operational steps: for each phase-space point $\alpha$, prepare (drive) the CNT state, apply $D(\alpha)$ (implemented by a calibrated resonant force pulse on the CNT), apply the CPB Ramsey-parity sequence with interaction time $t_\pi$, and read out the CPB. Repeating this for many $\alpha$ reconstructs $W(\alpha)$.

\paragraph{(d) Timescale requirements and backaction considerations.}
High-contrast mapping requires the controlled-phase gate time $t_\pi=\pi/\chi$ to be long compared to a mechanical period (so that the dispersive/RWA picture is meaningful) but short compared to relevant decoherence times. A conservative hierarchy is
\begin{equation}
\frac{2\pi}{\omega_0}\ll t_\pi \ll T_{2q},\,T_{1m},
\end{equation}
where $T_{2,q}$ is the CPB dephasing time during the Ramsey interval and $T_{2,m}$ is the mechanical coherence time relevant to the protocol. In addition, dispersive operation requires $|\Delta_q|\gg g_m$ to suppress real energy exchange and limit measurement backaction primarily to dephasing (photon/charge-noise induced). The rf-SET readout itself introduces backaction; however, in the standard staged protocol the rf-SET is effectively ``off'' (or biased to minimize backaction) during coherent evolution and is activated only during the final readout window, exactly as discussed in Ref.~\cite{blencowe_quantum_2004}.

\paragraph{(e) Summary of the approach.} Starting from the Jaynes--Cummings Hamiltonian and then taking the dispersive limit we obtain the number-dependent phase accumulation needed for parity mapping. In the the present section we clarified why this is physically well motivated: (i) the underlying coupling arises generically from gate-charge modulation by a mechanical electrode [Eq.~\eqref{SM:longitudinal}], (ii) near the charge degeneracy point the coupling becomes effectively transverse and reduces to the Jaynes--Cummings form under the RWA [Eq.~\eqref{SM:JC}], and (iii) in the large-detuning regime the dispersive interaction follows [Eq.~\eqref{SM:dispersive}], enabling a controlled $\pi$-per-phonon phase in time $t_\pi$ [Eq.~\eqref{SM:tpi}]. Finally, we note that mechanical implementations of dispersive couplings and superconducting-qubit-based readout have been pursued in multiple superconducting electromechanical platforms (CPB/transmon coupled to mechanical resonators), supporting the practicality of the measurement pathway assumed here \cite{lahaye_nanomechanical_2009,pirkkalainen_cavity_2015,lee_strong_2023}. The specific rf-SET-based CPB charge readout and pulse-sequence logic for transferring mechanical information to CPB populations is described explicitly in Ref.~\cite{blencowe_quantum_2004}.

\section{Estimates of the AFM--CNT interaction and backaction constraints}

The AFM apparatus can affect the CNT dynamics through two main channels:

(i) \emph{Position (gap) noise:} low-frequency fluctuations $\delta z(t)$ of the tip--CNT spacing modulate the static force gradient and thus the CNT resonance parameters. To leading order this appears as classical frequency noise,
\begin{equation}
\omega_{0}\ \rightarrow\ \omega_{0}+\delta\omega_{0}(t),
\qquad
\delta\omega_{0}(t)\simeq \left(\frac{\partial \omega_{0}}{\partial z}\right)\delta z(t),
\end{equation}
which contributes to dephasing and is naturally incorporated as an additional contribution to $\Gamma_2$ in the TLS master-equation description. The parameter regime required for the AFM-induced contribution to remain subdominant is that the induced frequency fluctuations are smaller than the intrinsic linewidth (including dephasing) and do not compromise the anharmonic selectivity, consistent with the TLS conditions in Eq. (2) in the main text. This is precisely the motivation for the ``tip-position noise'' discussion in the main text.

(ii) \emph{Voltage noise and force-noise at $\omega_0$:} fluctuations $\delta V(t)$ on the tip bias generate force noise $\delta F(t)$ through Eq. (26) in the main text. Near resonance, force noise at $\omega_0$ drives unwanted excitation and adds to the effective heating rate, while low-frequency voltage noise contributes primarily to dephasing through slow fluctuations of the electrostatic softening. In practice, both effects can be suppressed by (a) operating at modest $V_{\mathrm{dc}}$ so that $F_{\omega_d}\propto V_{\mathrm{dc}}V_{\mathrm{ac}}$ is set mainly by a controlled $V_{\mathrm{ac}}$, and (b) filtering/thermalizing the bias lines, as is standard in cryogenic electromechanics.

\paragraph*{AFM coupling to a thermal bath.} Our protocol does not require the cantilever to oscillate at $\omega_0$. The cantilever and positioning stack are used for quasi-static placement, while the resonant actuation is electrical. Consequently, the cantilever's coupling to its thermal bath enters primarily through \emph{low-frequency} displacement noise $\delta z(t)$ (treated above) rather than as a resonant dynamical element at $\omega_0$. Moreover, if the cantilever is thermalized to dilution-refrigerator temperatures, its \emph{Brownian} displacement scale is extremely small. A simple equipartition estimate gives an rms thermal displacement
\begin{equation}
x_{\mathrm{th,rms}}\simeq \sqrt{\frac{k_B T}{k_c}},
\end{equation}
so that at $T=10$~mK and $k_c\sim 1~\mathrm{N/m}$ one has $x_{\mathrm{th,rms}}\sim 1~\mathrm{pm}$, which is negligible compared to working separations $z_0$ of tens of nm. In this cryogenic, stiff-cantilever regime, thermal motion of the AFM mechanics does not by itself preclude stable approach. The remaining requirement is that \emph{instrumental} vibration and drift be sufficiently reduced so that the induced frequency noise and amplitude noise are small compared to the intrinsic linewidth and the control bandwidth (Eq. (2)in the main text). In Sec.~VI in the main text we therefore focus on experimentally realistic displacement-noise levels and force gradients to delineate the operating window in which AFM-induced dephasing/heating is subdominant.

Finally, we emphasize that the broader concept of applying time-dependent gradient forces via nearby electrodes/tips is well established in nanomechanical-qubit architectures, where static and rf potentials are explicitly used to tune the mechanical spectrum and to implement qubit rotations \cite{rips_hartmann_prl_2013}.

\paragraph{van der Waals interactions.} van der Waals forces provide an additional \emph{static} attractive background that can be accounted for in setting a safe working distance. Approximating the tip apex as a sphere of radius $R$ above a conductor gives the standard estimate \cite{giessibl_advances_2003}:
\begin{equation}
F_{\mathrm{vdW}}\simeq \frac{A_H R}{6 z_0^2},
\qquad
\frac{dF_{\mathrm{vdW}}}{dz}\simeq \frac{A_H R}{3 z_0^3},
\label{eq:Fvdw}
\end{equation}
with Hamaker constant $A_H\sim 10^{-19}$~J. For representative values $R\sim 20$~nm and $z_0\sim 50$~nm one finds $F_{\mathrm{vdW}}\sim 10^{-13}$~N (sub-pN) and $|dF_{\mathrm{vdW}}/dz|\sim 10^{-6}$~N/m, i.e.\ a background force that is ~ 2 orders of magnitude larger than the desired \emph{AC} drive amplitude but with a force gradient that remains far below typical cantilever stiffnesses ($k_c\sim 0.1$--$10$~N/m). In this regime, ``jump-to-contact'' instabilities (which occur when the attractive force gradient exceeds the mechanical restoring stiffness) are avoided \cite{giessibl_advances_2003}. Operationally, the static attraction is absorbed into a shifted equilibrium position and (if needed) compensated by adjusting $V_{\mathrm{dc}}$ so that only the small resonant component $F_{\omega_0}$ is used for control.

\section{Anharmonicity estimates, sideband cooling, and validity of the TLS approximation}
\label{SM:K_and_TLS_validity}

This section provides the standard relation between the classical Duffing nonlinearity and the quantum Kerr coefficient $K$, as well as estimates of anharmonicity $K$ at cryogenic temperatures, cooling strategies, a quantitative operating window for which the two-level truncation is controlled, and leakage estimates.

\textit{From Duffing nonlinearity to an effective Kerr coefficient.}
A single-mode flexural coordinate $q(t)$ of a doubly clamped beam with weak geometric nonlinearity is commonly modeled (classically) by the Duffing equation \cite{lifshitz_cross_nonlinear_2009}
\begin{equation}
m_{\mathrm{eff}}\ddot q + m_{\mathrm{eff}}\gamma_m \dot q + m_{\mathrm{eff}}\omega_0^2 q + \alpha\,q^3 = F(t),
\label{SM:duffing_classical}
\end{equation}
where $\alpha$ is the cubic restoring-force coefficient (equivalently, a quartic correction to the potential energy). The corresponding Hamiltonian is
\begin{equation}
H = \frac{p^2}{2m_{\mathrm{eff}}} + \frac{1}{2}m_{\mathrm{eff}}\omega_0^2 q^2 + \frac{\alpha}{4}\,q^4 - F(t)\,q.
\label{SM:Hq}
\end{equation}
Quantizing via $q = x_{\mathrm{zpf}}(a+a^\dagger)$ with
$x_{\mathrm{zpf}}=\sqrt{\hbar/(2m_{\mathrm{eff}}\omega_0)}$ and applying the rotating-wave approximation to the quartic term yields the standard weak-nonlinearity form \cite{semiao_kerr_2009, rips_nonlinear_2014}
\begin{equation}
H \simeq \hbar\omega_0\,a^\dagger a + \frac{\hbar K}{2}\,a^\dagger a^\dagger a a \;+\; \cdots,
\label{SM:Hkerr}
\end{equation}
with the leading-order relation
\begin{equation}
K \simeq \frac{3\alpha\,x_{\mathrm{zpf}}^{4}}{2\hbar}.
\label{SM:K_alpha_relation}
\end{equation}
(Up to convention-dependent numerical factors that depend on the precise definition of $\alpha$ and on the RWA bookkeeping, Eq.~\eqref{SM:K_alpha_relation} captures the standard scaling $K\propto \alpha x_{\mathrm{zpf}}^4/\hbar$.) Equation~\eqref{SM:Hkerr} implies a level-dependent transition frequency $\omega_{n,n+1}\approx \omega_0 + nK$, so that the anharmonic separation is
\begin{equation}
\omega_{21}-\omega_{10}\approx K.
\label{SM:anharm_sep}
\end{equation}

A key practical point is that electrostatic softening enhances the nonlinearity per phonon because $x_{\mathrm{zpf}}\propto \omega_0^{-1/2}$ and therefore $K\propto x_{\mathrm{zpf}}^4\propto \omega_0^{-2}$ for fixed underlying elastic nonlinearity \cite{rips_hartmann_prl_2013}. This scaling underlies the numerical estimates below.

\textit{Active cooling.} As discussed in the main at $T=10 \ \mathrm{mK}$, GHz-frequency CNT modes are near the ground state, while MHz-scale modes have substantial thermal occupation and therefore require active cooling (sideband or feedback) to reach \(\bar n_{\mathrm{th}}\ll 1\). Using the representative frequencies in Table I (main text), one finds \(\bar n_{\mathrm{th}}\simeq 6\times 10^{-12}\) for \(\omega_0/2\pi=5.37~\mathrm{GHz}\), but \(\bar n_{\mathrm{th}}\simeq 0.5\) and \(3.4\) for \(\omega_0/2\pi=221~\mathrm{MHz}\) and \(54~\mathrm{MHz}\), respectively. Accordingly, short CNTs (GHz-class devices) can directly operate in the low-\(\bar n_{\mathrm{th}}\) regime at base temperature, enabling high-contrast state preparation and parity-based Wigner tomography without additional cooling overhead. Longer CNTs (10–100~MHz) devices have to implement active cooling via either (a) sideband cooling in a cavity or CPB-assisted readout configuration and/or (b) measurement-based feedback cooling using the AFM/cavity transducer as the continuous monitor, as achieved in cavity optomechanics, QED and superconducting circuit - based experiments  \cite{aspelmeyer_cavity_2014,cui_feedback_2013,riste_initialization_2012}. We note that in all cases, \(\bar n_{\mathrm{th}}\) can be experimentally calibrated (and tracked during gate sequences) from the reconstructed \(W(\alpha)\) and from independent ringdown spectra, so that finite-temperature effects enter the GKSL modeling as measured, not assumed.

\textit{Benchmark value for $K$, scaling, and leakage estimate.}
The sufficient conditions used in the manuscript,
\begin{equation}
|K| \gg \Gamma_2,
\qquad
|K| \gg \Omega_R,
\label{SM:tls_cond}
\end{equation}
have transparent physical meaning: the first ensures that adjacent transitions remain spectrally resolvable despite damping (and any additional dephasing folded into an effective linewidth), while the second ensures the control field does not significantly address off-resonant higher transitions during gate operations \cite{rips_hartmann_prl_2013,wang_method_2016,sarma_tunable_2018}. 
Reference~\cite{rips_hartmann_prl_2013} provides a concrete estimate for nanotubes in the softening-enhanced regime: by tuning a CNT mechanical mode to
\begin{equation}
\omega_{\mathrm{ref}}/2\pi \simeq {26.6}\ \mathrm{MHz},
\qquad
K_{\mathrm{ref}}/2\pi \equiv (\omega_{21}-\omega_{10})/2\pi \simeq {2.71}\ \mathrm{MHz},
\label{SM:benchmark}
\end{equation}
i.e.an effective \ a MHz-scale anharmonic separation in a realistic CNT geometry. 
In this regime, the two-level selectivity conditions are readily satisfied: \( |K|\gg \Gamma_2\) for \(Q\sim 10^4\), since $\Gamma_2\simeq\Gamma_1/2= \omega_0/(2Q)\sim 10-100~\mathrm{kHz}$, and \( |K|\gg \Omega_R\) is ensured by choosing control bandwidths below the anharmonic separation (or equivalently, using longer and spectrally narrower pulses). Moreover, the scaling relation $K\propto \omega_0^{-2}$ (see above), illustrates a central tradeoff: higher-frequency devices are thermally favorable at {10}{mK} but typically have smaller intrinsic anharmonicity per phonon, while softened MHz-frequency devices can reach $K$ in the {0.1--10 MHz} range, enabling controlled two-level operation provided the mode is actively cooled \cite{rips_hartmann_prl_2013}.

For control, we have $\Omega_R = \pi/(2t_{\pi/2})$ for a $\pi/2$ pulse of duration $t_{\pi/2}$. The choice \(t_{\pi/2}={0.1}\ {\mu}s\) used for Table I corresponds to $\Omega_R/2\pi\simeq{2.5}{\ \mathrm{MHz}}$. In the MHz-anharmonicity regime it is therefore natural to choose longer pulses, e.g.\ $t_{\pi/2}={1}{{\ \mathrm{\mu}s}}$ giving $\Omega_R/2\pi\simeq{0.25} {\ \mathrm{MHz}}$, which yields $|K|/\Omega_R\gtrsim 10$ for $K/2\pi\sim{2.7}{\ \mathrm{MHz}}$ and makes leakage parametrically small.
A simple perturbative estimate of leakage to level $|2\rangle$ during resonant driving of $|0\rangle\!\leftrightarrow\!|1\rangle$ is obtained by treating the unwanted $|1\rangle\!\to\!|2\rangle$ coupling as off-resonant by detuning $\Delta_{12}\approx K$; the resulting excited-state amplitude is suppressed as $\sim \Omega_R/\Delta_{12}$, giving a leakage probability scaling \cite{motzoi_simple_2009}
\begin{equation}
P_{\mathrm{leak}} \sim \left(\frac{\Omega_R}{K}\right)^2,
\label{SM:leak_scaling}
\end{equation}
up to order-unity factors depending on pulse shape. Equation~\eqref{SM:leak_scaling} makes explicit why the separation of scales in Eq.~\eqref{SM:tls_cond} is the appropriate quantitative criterion for confining the driven dynamics to $\{|0\rangle,|1\rangle\}$.

\textit{In summary,} the combined requirements for the mechanical-qubit regime are therefore:
\begin{enumerate}
\item \textbf{Initialization:} $\bar n_{\mathrm{th}}\ll 1$ (achieved either by large $\omega_0$ at fixed $T$ or by active cooling).
\item \textbf{Spectral selectivity:} $|K|\gg \Gamma_2$ and $|K|\gg \Omega_R$ .
\end{enumerate}
Our analysis shows that these conditions are mutually compatible in the electrostatically softened CNT regime, which yields MHz-scale anharmonicity at tens of MHz frequency, together with active cooling to reach $\bar n_{\mathrm{th}}\ll 1$ and pulse durations chosen such that $\Omega_R\ll |K|$.

\end{document}